
\input harvmac
\def\half{{1\over 2}}

\def\sF{{\cal F}}
\def\sG{{\cal G}}
\def\sP{{\cal P}}
\def\ker{{\rm Ker}}
\def\cok{{\rm Cok}}
\def\ext{{\rm Ext}}
\def\hom{{\rm Hom}}
\def\zbar{\overline{z}}
\def\vac{\vert 0\rangle}
\def\O{{\cal O}}
\def\oo{\overline}
\def\cA{{\cal A}}
\def\cB{{\cal B}}

%
%
\def\eqnn#1{\xdef #1{(\secsym\the\meqno)}\writedef{#1\leftbracket#1}%
\global\advance\meqno by1\wrlabeL#1}
\def\eqna#1{\xdef #1##1{\hbox{$(\secsym\the\meqno##1)$}}
\writedef{#1\numbersign1\leftbracket#1{\numbersign1}}%
\global\advance\meqno by1\wrlabeL{#1$\{\}$}}
\def\eqn#1#2{\xdef #1{(\secsym\the\meqno)}\writedef{#1\leftbracket#1}%
\global\advance\meqno by1$$#2\eqno#1\eqlabeL#1$$}

\nref\rMax{M. Kontsevich, private communication.}
\nref\rCVNEW{S. Cecotti and C. Vafa, ``On Classification of $N = 2$
Supersymmetric Theories,''  Harvard and SISSA preprints HUTP-92/A064 and
SISSA-203/92/EP.}
\nref\rCVTAF{S. Cecotti and C. Vafa, Nucl. Phys {\bf B367}
(1991) 3543.}
\nref\rLVW{W. Lerche, C. Vafa, and N. Warner, Nucl. Phys. {\bf B324}
(1989) 427.}
\nref\rHelices{A. N. Rudakov et al., {\sl Helices and Vector Bundles:
Seminaire Rudakov,} London Mathematical Society Lecture Note Series 148,
Cambridge University Press, Cambridge 1990.}
\nref\rW{E. Witten, ``Mirror Manifolds and Topological Field
Theory," Institute for Advanced Studies preprint IASSNS-HEP 91/83.}
\nref\rMorse{E. Witten, J. Diff. Geom. {\bf 17} (1982) 661.}
\nref\rCecotti{S. Cecotti, L Girardello, and A. Pasquinucci, Intl. J. Mod.
Phys. {\bf A6} (1991) 2427.}
\nref\rTop{E. Witten, Commun. Math. Phys. {\bf 118} (1988) 411; S. Cecotti,
Intl. J. Mod. Phys. {\bf A6} (1991) 1749.}
\nref\rNEWSUSY{S. Cecotti, P. Fendley, K. Intriligator, and C. Vafa, Nucl.
Phys. {\bf B 386} (1992) 405.}
\nref\rPIII{A. R. Its and V. Yu. Novokshenov, {\sl The Isomonodromic
Deformation Method in the Theory of Painlev\'e Equations,} Lecture Notes
in Mathematics 1191, Springer-Verlag, New York 1986; B.M. McCoy, C.A.
Tracy, and T.T. Wu, J. Math Phys. {\bf 18} (1977) 1058.}
\nref\rKen{K. Intriligator, private communication.}
\nref\rWCPN{E. Witten, Nucl. Phys. {\bf B149} (1979) 285.}
\nref\rCVMO{S. Cecotti and C. Vafa, Mod. Phys. Lett. {\bf A7} (1992) 1715.}
\nref\rZDD{E. Zaslow, Nucl. Phys. {\bf B415} (1994) 155.}
\nref\rHV{S. Hamidi and C. Vafa, Nucl. Phys. {\bf B279} (1989) 465.}
\nref\rZAZ{E. Zaslow, Comm. Math. Phys. {\bf 156} (1993) 301.}
\nref\rSLOD{J. McKay, {\sl Cartan Matrices, Finite Groups of Quaternions, and
Kleinian Singularities,} Proc. Am. Math. Soc. (1981) 153; P. Slodowy, {\sl
Simple Singularities and Simple Algebraic Groups,} Lecture Notes in Mathematics
815, Springer-Verlag, New York 1980.}
\nref\rGH{P. Griffiths and J. Harris, {\sl Principles of Algebraic Geometry.}}
\nref\rGor{A. L. Gorodentsev, Math. USSR Izv. {\bf 33} (1989) 67.}
\nref\rKapranov{M. M. Kapranov, Invent. Math. {\bf 92} (1988) 479; M. M.
Kapranov, Math. USSR Izv. {\bf 24} (1985) 183.}
\nref\rKappriv{M. M. Kapranov, private communication.}
\nref\rBK{A. I. Bondal and M. M. Kapranov, Math. USSR Izv. {\bf 35} (1990)
519.}
\nref\rPol{A. E. Polischuk, private communication.}
\nref\rBondal{A. I. Bondal, Math. USSR Izv. {\bf 34} (1990) 23.}
\nref\rRud{A. N. Rudakov, Math USSR Izv. {\bf 32} (1989) 99.}
\nref\rPot{L. Alvarez-Gaum\'e and D. Z. Freedman, Commun.
Math. Phys. {\bf 91} (1983) 87.}
\nref\rCL{J. B. Carrell and D. I. Lieberman, Inv. Math. {\bf 21} (1973) 303.}
\nref\rRaj{R. Rajaraman, {\sl Solitons and Instantons,} Elsevier Science
Publishers, Amsterdam, 1987.}
\nref\rBat{V. V. Batyrev, ``Quantum Cohomology Rings of Toric Manifolds,'' MSRI
preprint, 1993.}
\nref\rWLH{W. A. Leaf-Herrmann, Harvard preprint HUTP-91/AO61.}
\nref\rNogin{D. Yu. Nogin, {\sl Algebraic Geometry,} Lecture Notes in
Mathematics 1479, Springer-Verlag, New York 1991.}

\nfig\projdiag{$P$ is projective when this diagram holds.}
\nfig\quivdiag{An ordered quiver with three veritices $p_i,$ three arrows $g_i$
connecting $p_1$ to $p_2$ and three arrows $f_i$ connecting $p_2$ to $p_3.$}
\nfig\toricdiag{Diagram representing the toric variety $\widetilde{\bf P}^2$
with four vectors in a plane:  $\vec{v}_1 = (0,-1),$ $\vec{v}_2 = (1,0),$
$\vec{v}_3 = (1,0),$ $\vec{v}_4 = (-1,1).$}

\Title{HUTP-94/A027}{\vbox{\centerline{Solitons and Helices:}
\vskip2pt\centerline{The Search for a Math-Physics Bridge}}}

\centerline{Eric Zaslow\footnote{$^*$}
{Supported in part by Fannie and John Hertz Foundation.}}

\bigskip\centerline{Lyman Laboratory of Physics}
\centerline{Harvard University}\centerline{Cambridge, MA 02138}

\vskip .3in

We present evidence for an undiscovered link between N=2 supersymmetric quantum
field theories and the mathematical theory of helices of coherent sheaves.  We
give a thorough review for nonspecialists of both the mathematics and physics
involved, and invite the reader to take up the search for this elusive
connection.

\Date{8/94}

\newsec{Introduction}
Last year, Kontsevich noticed a similarity between the work of Cecotti and Vafa
on classifying two-dimensional N=2 supersymmetric field theories and the work
of some
algebraic geometers in Moscow \rMax.  In the independent and seemingly
unrelated work
of physicists and mathematicians, similar structures emerged.  Both had found
quasi-unipotent matrices satisfying certain Diophantine conditions, which
supported the action of the braid group.  Were they the same?

Behind this question lie a potential relationship between disparate fields and
the opportunity for string theory and its offshoots to once again bring
mathematicians and physicists together.  Unfortunately, my search for this
bridge was somewhat in vain.  I cannot tout complete success; instead I offer
an amalgam of evidence and observations supporting this conjecture, along with
various approaches used in trying to find this elusive link.  These diverse
techniques span a breadth of physics and mathematics.  This paper is intended
to give a thorough treatment while remaining somewhat self-contained, perhaps
at the expense of brevity.

The physics is the theory of classifying two-dimensional N=2 supersymmetric
field theories \rCVNEW\ and is closely related to topological-anti-topological
($tt^*$) fusion \rCVTAF.  The idea for classification was to obtain information
about the number of vacua and solitons between them in the infrared limit.
Given a massive N=2 theory (we will always consider two-dimensional theories),
one can consider the whole renormalization group trajectory -- its infrared and
ultraviolet limits.  In the conformal, or ultraviolet, limit, the (universality
class of the) theory can be partially classified by the structure of the chiral
primary ring \rLVW.  In particular one can compute the number of ring elements
and their $U(1)$ charges.  In the infrared limit, there is no superconformal
symmetry; the excitations above the vacua become infinitely massive and we can
investigate tunneling between vacua.  These amplitudes will reveal the numbers
of solitons connecting different vacua.  These numbers, too, help classify the
theory.  In fact, the $U(1)$ charges of the theory at its conformal limit may
be derived from this information.  Our interest is in the topological sigma
model associated to a K\"ahler target space.  The physical picture is detailed
in section two.

The mathematics involved regards a branch of study which has been developed in
the past decade primarily at Moscow University \rHelices.  The theory is that
of collections of coherent sheaves called helices.  The theory arose from the
study of vector bundles over low dimensional projective spaces.  Unfortunately,
the math is almost as new as the physics, and few results are known rigorously.
 Helices are collections of sheaves over some complex base manifold obeying a
sort of upper-triangularity condition (on the Euler characteristic between
sheaves).  These collections transform under mutations defining an action of
the braid group on a finite collection of sheaves -- the {\sl foundation} --
from which the helix is determined.    A bilinear form over the foundation is
defined; in several examples, we see that it is precisely the matrix derived
from the physical theory on the same base manifold.  Section three is composed
of a detailed exploration of this mathematical subject.

Even without an explicit connection between the two disciplines, it may be
possible to prove some sort of categorical equivalence between certain classes
of N=2 quantum field theories and foundations of helices.  Such a description,
while interesting, would not be as enlightening.  For example, we have noticed
that the matrices associated to topological sigma models over a given manifold
correspond to the matrices associated with bundles over the same space.  A
categorical equivalence could not guarantee that the base space of the vector
bundles and the sigma model should be in correspondence.  We will discuss this
further, along with several techniques which may be useful in finding a bridge,
in the section four.

A note on point of view: In the section labeled ``The Physics," the first
person ``we" is taken to mean ``we mathematicians."  In the section on
mathematics, it means ``we physicists.''  This schizophrenic viewpoint reflects
only the author's natural identification with those who feel inexpert.
\vfill
\eject

\newsec{The Physics}
\subsec{Overview}
The physics we will discuss involves the realm of two-dimensional quantum field
theories with two independent supersymmetry charges.  These theories have many
interesting properties which relate to various fields of mathematics including
de Rham and Hodge cohomologies \rW; Morse theory \rMorse; singularities and
Picard Lefschetz theory \rCVNEW,\rCecotti; variations of Hodge structure.  The
latter two are more closely related to our area of investigation, and are
particular to the N=2 case (these structures are absent in N=1 supersymmetric
models).  Specifically, we will be working in the topological sector of such
theories \rTop.  This is interpreted as follows.  There exist a certain set of
correlations functions in these theories which are independent of the positions
of the fields on the two-dimensional surface.  If we restrict the set of fields
and correlation functions to those which have this property, we can use the N=2
theory as a means of creating a ``topological theory.''  Typically, the space
of topological field theories is composed of finite-dimensional components.
This space can be thought of as the parameter space of the quantum field
theory, where only the topologically relevant parameters -- i.e. those which
when perturbed change the topological correlation functions -- are considered.

Thus, our spaces will be finite-dimensional spaces, each point of which will
represent a topological field theory.  Let us fix a component ${\cal M}$ of
this space, and label its points by $t.$  The fields $\phi_i$ in these theories
create states $\vert i\rangle$ which comprise a Hilbert space, ${\cal H}_{t}.$
The fields (and hence states) will typically transform amongst themselves under
a perturbation.  They thus define a vector bundle over moduli space, with a
natural connection we will determine presently through field theoretic methods.
 The fields define a commutative associative algebra:  $\phi_i\times\phi_j =
C_{ij}{}^k\phi_k.$  The correlation functions, being independent of position,
can only depend on the types of fields and on the point ${t}.$  We will be
concerned with the variation of these parameters with the moduli.

As we will show, the Hilbert spaces can be thought of as spaces of $vacua,$
i.e. states of zero energy.  If we imagine a potential, the vacua lie at the
minima, which we will take to be discrete and labeled $x_n.$  If space is the
real line, a field configuration $\phi(x)$ satisfying $\phi(\infty) = x_a$ and
$\phi(-\infty) = x_b$ is said to be in the $ab$ soliton sector.  A minimal
energy configuration is a soliton.  The situation is decidedly more difficult
to interpret in field theories without potentials.  After our discussion of
topological field theories, we will relate some of the quantities discussed
above to the numbers of solitons in the theory.  The sigma models are theories
defined for a given Kahler manifold and Kahler form, ${\bf k.}$  The moduli
space of theories for a given manifold will be the Kahler cone.  To each such
$M$ we will derive a matrix encoding the soliton numbers, which has several
interesting properties.  It transforms under the action of a braid group and is
quasi-unipotent.  We will liken this to a similar matrix derived through the
theory of helices.

\subsec{Topological Field Theory}

Topological field theories are models in which correlation functions do not
depend on the positions of the operators involved.  They therefore depend only
on the type of operators involved, and on topological properties of the space
of field configurations.  In the case of topological sigma models, in which the
quantum fields are maps to some target manifold, the topology of the target
manifold becomes crucial.  Such a theory can be defined given any N=2 quantum
field theory.  Topological theories constructed in this manner will be studied
here.

Let us describe the twisting procedure which yields a topological theory from
an N=2 theory.  To do so, let us first consider a theory defined on an infinite
flat cylinder.  A quantum theory with N=2 supersymmetry is invariant under the
N=2 superalgebra.  This algebra contains two fermionic generators $Q^1, Q^2,$
as well as bosonic generators, which mix non-trivially.  There is also an
$SO(2)$ automorphism of this algebra rotating the $Q$'s.  We usually write
$Q^\pm = \half(Q^1 \mp iQ^2),$ where the sign denotes the charge under the
$SO(2)$ generator, $J.$  Further, since the supercharges are spinorial (they
give spinors from bosons), their components have a chirality in two-dimensions.
 This gives us four charges:  $Q^+_L, Q^+_R, Q^-_L, Q^-_R.$  The algebra
contains the two dimensional Lorentz group as well and reads:
\eqn\ntwosualg{\matrix{\eqalign{\{Q^+_L,Q^-_L\} &= 2H_L,
\cr [L,Q^{\pm}_L] &= \half Q^\pm_L,\cr [L,H_L] &= H_L,\cr
[J,Q^\pm_L] &= \pm Q^\pm_L, \cr} & \qquad\eqalign{\{{Q}^+_R,{Q}^-_R\} &= 2H_R,
\cr [L,{Q}^{\pm}_R] &= \half {Q}^\pm_R,\cr [L,H_R] &= -H_R,\cr
[J,{Q}^\pm_R] &= \pm {Q}^\pm_R, \cr}}}
with all other (anti-)commutators vanishing.\footnote{$^*$}{This algebra is
modified in soliton sectors.  There, it includes central terms which yield the
Bogolmonyi bound.}  Here $L$ is the generator of the Euclidean rotation group
$SO(2)$ and $J$ is the generator of the $SO(2)$ rotation mixing the $Q$'s.
Also, $H_{L,R} = H\pm P.$

The topological theory is defined cohomologically by constructing a boundary
operator from the $Q'$s.  Let us define, then,
$$Q_+ \equiv Q^+_L + Q^+_R.$$
Note that
$$(Q_+)^2 = 0.$$
We use this operator to define cohomology classes, reducing the space of states
to a finite-dimensional Hilbert space:
$${\cal H} \equiv H^*(Q) = {\{ \vert\psi\rangle \> :\> Q_+\vert\psi\rangle =
0\}\over
\{ \vert\psi\rangle \> :\> \vert\psi\rangle = Q_+\vert\Lambda\rangle\} }.$$
Similarly, the fields are defined modulo commutators with $Q_+.$  Topological
invariance follows from the fact that derivatives with respect to $z$ and
$\zbar$ are represented by the action of $H_L$ and $H_R,$ respectively, on the
fields.  Since both $H_L$ and $H_R$ are exact ($H_{L,R} =
\half\{Q_+,Q^-_{L,R}\}),$ all correlation functions between topological states
wil be invariant under infinitesimal variations of the positions.

We would like to extend this analysis to arbitrary Riemann surfaces.  What
prevents us from doing so now is that $Q_+$ is made up not of scalars but of
pieces of spinors.  When our surface was a flat torus with trivial spin
connection, the separate components of $Q_+$ and $Q_-$ were globally defined.
These will not be globally defined on a general Riemann surface, and so what
was cohomologically trivial in one set of coordinates may be nontrivial in
another.  To remedy this, we simply {\sl declare} $Q_+$ to be a scalar.  That
is, we can redefine the spin of the fermions by adding a background gauge field
proportional to the spin connection.

The topological fields form a ring, just as de Rham cohomology elements form a
ring.  The products are well-defined, since we can note $(\phi_1 + [Q_+,\Lambda
])\phi_2 = \phi_1\phi_2 + [Q_+,\Lambda\phi_2] \equiv \phi_1\phi_2,$ which
follows from $[Q_+,\phi_2] = 0.$  Let us choose a set of generators $\phi_i$
for the topological field space.  The operator product can be captured through
the structure constants $C_{ij}{}^k$ by writing
$$\phi_i\times\phi_j = C_{ij}{}^k\phi_k.$$
The field space is in one-to-one correspondence with the Hilbert space by the
relation
$$\phi_i\vac \equiv \vert i\rangle.$$
In the above we have used the unique vacuum $\vac$ from the N=2 quantum field
theory.\footnote{$^*$}{This vacuum is the unique vacuum of the Neveu-Schwartz
sector.}
The correlators are then all given by the ring coefficients $C_{ij}{}^k$ and
the two-point function \eqn\topmet{\eta_{ij} = \langle i\vert j\rangle.}
Note that we don't take the adjoint state $\langle \oo{i}\vert$ in forming the
topological metric.  In \topmet, $\langle i\vert$ obeys
$$\langle i\vert Q_+ = 0,$$
which insures topological invariance of the correlation functions.  We note
here that in the particular case where $\vert i\rangle$ is a ground state, and
therefore annihilated by $Q_+$ and $Q_-,$ we can take the regular adjoint and
the correlation functions will still be topological.  The discovery of such
independence in correlators was made in particular models several years before
topological field theory was systematically treated.

The analogy with de Rham cohomology can be extended to Hodge cohomolgy.  We can
interpret the Hamiltonian as the Laplace operator, with $Q_+$ and $Q_-$ serving
as the $\partial$ and $\oo{\partial}$ operators.  Then, as with Hodge
decomposition, we have the following statement.  Every $Q_+$ cohomology class
has a unique harmonic representative, i.e. a unique representative annihilated
by $Q_-.$  Noting that zero energy states are annihilated by $Q_+$ and $Q_-$ we
have the equivalence of several vector spaces:\footnote{$^{**}$}{We see
$\langle\psi\vert H\vert\psi\rangle = 0 \Rightarrow\langle\psi\vert(Q_-Q_+ +
Q_+Q_-)\vert\psi\rangle = 0 \Rightarrow \| Q_+\vert\psi\rangle\|^2 + \|
Q_-\vert\psi\rangle\|^2 = 0,$ since $(Q_-)^\dagger = Q_+.$  Therefore both
terms, being positive definite in a unitary theory, are zero separately.}
\eqn\harmony{Q_+ \hbox{ cohomology} \longleftrightarrow \hbox{Vacua}
\longleftrightarrow Q_- \hbox{ cohomology}.}  The second equivalence in
\harmony\ is made simply by interchanging the r\^oles of $Q_+$ and $Q_-.$

The simple observation \harmony\ will provide us with a rich source for
exploration.  Specifically, we will ask how the isomorphism between the $Q_-$
and $Q_+$ cohomologies varies over the space of topological field theories.

To illustrate the structure of topological theories and provide us with our
main object of study, we briefly discuss the structure of the chiral ring for
the topological sigma models.  By sigma model, we mean a quantum field theory
in which the bosonic variables are maps (from a two-dimensional surface) to a
target manifold.  In the N=1 and N=2 supersymmetric theories, the fermionic
structures mimic the forms of de Rham and Hodge cohomology.  The action takes
the form
\eqn\action{
S = 2t\int_\Sigma \!d^2\!z\, {1\over 2} g_{IJ}
\partial_z\phi^I\partial_{\overline{z}} \phi^J +
i\psi_{-}^{\overline{i}} D_{z} \psi_-^i g_{\overline{i} i} +
i\psi_+^{\overline{i}} D_{\overline{z}} \psi_+^i g_{\overline{i} i}
+ R_{i\overline{i}j\overline{j}} \psi_+^i \psi_+^{\overline{i}}
 \psi_-^j \psi_-^{\overline{j}}
.}
Here $\Sigma$ represents the Riemann surface, which, for our purpose will
always be of genus zero, $g_{IJ}$ and $R_{i\overline{i}j\overline{j}}$
are respectively the metric and Riemann tensor of the target space.  $D$
is the pull-back onto $\Sigma$ of the connection under the
map, $\Phi.$  The $N = 2$ structure implies a holomorphic $U(1)$
current, by
which we may twist the energy-momentum tensor.  That is, we can redefine spins
by adding a background gauge field equal to (one half) the spin connection.
Mathematically, this is
equivalent to redefining the bundles in which the fields live.  As we have
discussed, this will render the BRST charge $Q_+$ a scalar on the Riemann
surface, so that the theory is defined for any genus.
Specifically, we now take $\psi_+^i \in \Phi^{*}(T^{1,0})$ and
$\psi_-^{\overline{i}} \in \Phi^{*}(T^{0,1}).$
We put $\psi_+^{\overline{i}} \in \Omega^{1,0}
(\Sigma ; \Phi^{*}(T^{0,1}))$ and
$\psi_-^i \in
\Omega^{0,1}(\Sigma;\Phi^{*}(T^{1,0}));$ that is, they combine to form
a one-form on $\Sigma$
with values in the pull-back of the tangent space of $K:$ call
these components $\psi_z^{\overline{i}}$ and $\psi_{\overline{z}}^i$
respectively.

The important aspect of these theories are that the energy-momentum tensor is
$Q_+$-exact.  This allows us to rescale the two-dimensional metric $\delta
h_{\mu\nu} = \Lambda h_{\mu\nu}$
without affecting the correlators.  As $\Lambda\rightarrow\infty$ -- the
topological limit -- the only non-vanishing contribution to the path integral
is from the instanton configurations, or classical minima.  As the space of
instantons, ${\cal M},$ is disconnected, the computation reduces to a sum over
components of ${\cal M}.$  Supersymmetry ensures the cancellation of the
determinant from bosonic and fermionic oscillator modes.  The zero mode
integration yields just the number of instantons taking the insertion points to
the Poincar\'e dual forms representing the corresponding operators.  This is
how we derive the ring of observables.

\subsec{Topological-Anti-Topological Fusion and Classification of N=2 Theories}

We will be interested in the numbers of (Bogolmonyi saturating) solitons
connecting ground
states.  These numbers were used by Cecotti and Vafa in their
classification of N=2 superconformal theories with massive deformations
\rCVNEW.  The idea is that in the infrared limit, the two-point functions of
different vacua
(choosing an appropriate basis) obey
\eqn\tunnel{\langle\oo i\vert j\rangle \sim \delta_{ij} + {\rm tunneling\,\,
corrections}.}
The tunneling corrections indicate the presence of solitons, and will
depend on the size (Kahler class) of the manifold, or more generally the
couplings of the theory.  The tunneling corrections vanish in the
infinite volume (conformal) limit, but the asymptotic behavior will
indicate the number of solitons present (in a manner which will be made
explicit).  The dependence on the couplings is
described by the $tt^*$ equations of reference \rCVTAF.  We review this
technology below, then discuss the Diophantine constraints of classification.

In the previous section, we discussed how to make a topological field theory
given any N=2 theory by taking the $Q_+$ cohomology classes as states.
Alternatively, we could have defined a theory with the $Q_-$ cohomology.  We
can call this theory the ``anti-topological'' theory (it is still a topological
field theory).  In \harmony\ we noted that the spaces of states were
isomorphic.  They can be thought of as different bases for a finite dimensional
vector space.  This means that each anti-topological state $\vert\oo{a}\rangle$
can be expressed in terms of the topological states:
$\vert\oo{a}\rangle = \sum_{b} C_{b}\vert b\rangle,$
for some coefficients $C_{b}.$  More generally, we write
$$\langle \oo{a}\vert = \langle b\vert M^b{}_{\oo{a}},$$
with the sum over $b$ understood.  In this section we will describe how to
compute this change of basis, and its variation on theory space.  To do this,
let us examine the relationship between the topological and anti-topological
field theories.

The quantum field theory defines a metric on the topological Hilbert
space, $$g_{a\overline{b}} = \langle \overline{b} \vert a \rangle,$$
where we require the states to be ground states.  This metric thus fuses
the topological and anti-topological theory.  In fact, by connecting two
hemispherical regions along a common flat boundary, we can perform the
topological twist on one half and the anti-topological twist on the other
half.  The long cylindrical middle projects states to their ground
state representatives; flatness allows us to conjoin the
different
background metrics, used to make the topological twist,
where they vanish.  The resulting metric is the one described above, and
is independent of the representatives of the topological states.

The topological theory defines a symmetric topological metric, given by
intersections in an
appropriate moduli space of classical minima:
$$\eta_{ab} = \langle a\vert b\rangle$$ ($\eta_{\oo a\oo b} = \eta_{ab}^*$).
We note that $M = \eta^{-1}g$ and $(CPT)^2 = 1$ implies $MM^* = 1$ by relating
topological and anti-topological states.

These structures are defined for any N=2
theory, and become geometrical structures on the space of theories.
We can coordinatize this space by coupling constants $\{ t_i \}.$
The key observation for us is that correlation functions of neighboring
theories can be computed in terms of correlators of a given theory.  So let us
consider a theory described by an action $S(0)$ at some point $t=0$ in
parameter space.  We can parametrize a neighborhood of $0$ by perturbing by the
local operators $\phi_i.$  Thus, we write
\eqn\basept{S(t) = S(0) - \sum_i \left[ \int d^2\theta^- d^2z t_i\phi_i\> +\>
\hbox{h.c.}\right],}
where the perturbation is assumed to be small.  The correlations functions are
now $t-$dependent, but we can compute the variation given knowledge of the
theory at $t = 0.$
Indeed,
$$\del_i\langle\phi_1(z_1)...\phi_n(z_n)\rangle_{t=0} =
\langle\int\phi_i(z)d^2zd^2\theta^-\phi_1(z_1)...\phi_n(z_n)\rangle_{t=0},$$
where $\del_i \equiv \del/\del t_i$ and the subscript indicates that the
correlation functions are evaluated at $t=0.$
At each $t,$ we have a chiral ring, isomorphic to the Ramond ground states
of the theory.  We thus have a vector bundle -- the bundle of ground states --
with the metric given above
(now $t$-dependent).  A ground state, characterized by its $U(1)$ charge, is
then a section of this bundle;
its wave function is therefore $t$-dependent, and we can thus consider
the connection defined by
$$(A_i)_{a\overline{b}} = \langle \overline{b} \vert \del_i \vert a
\rangle.$$
Here we project out the change in $\vert a\rangle$ orthogonal to the ground
states.  The covariant derivative is then $D_i = \del_i - A_i.$  This
connection is defined so that
$$D_ig_{a\oo b} = 0,$$ which follows simply.

The equations of topological-anti-topological fusion, the $tt^*$
equations, describe the dependences of our geometrical constructions on
the couplings $t_i$ and $\oo t_{\oo i}.$  The equations may be expressed
covariantly, or in a particular choice of basis for the ground states
(gauge).  We will first show that the topological states (i.e. the $Q_+$
cohomology) constitute a ``holomorphic basis,'' in which anti-holomorphic
part of the connection vanishes:  $A_{\oo i}{}_a{}^b = 0.$  To prove this
we note that in the path-integral formalism, the state $\vert a\rangle =
\vert\phi_a\rangle$ is given by the path-integral over $S_R,$ the right
half of a sphere; so we have, from \basept:
$$\eqalign{\del_{\oo i}\vert a\rangle &=
\vert\int_{S_R}d^2\theta^+d^2z\oo\phi_{\oo i}(z)\phi_a\rangle\cr &=
Q^+\oo Q^+\vert\int_{S_R}d^2z\oo\phi_{\oo i}(z)\phi_a\rangle\cr}$$
(we adopt the convention $Q^+ = D^+$).
Then it is clear that the projection
to states obeying $\langle c\vert Q^+ = 0$ kills $A_{\oo i}:$
$$(A_{\oo i})_a{}^b =
\eta^{bc}\langle c\vert \del_{\oo i}\vert a\rangle = 0.$$
Clearly, we could also choose the anti-topological basis, which would
yield $A_i{}_{\oo a}{}^{\oo b} = 0.$  In the holomorphic basis, the
covariant constancy of the metric determines the connection:
$$0 = D_ig_{a\oo b} =
\del_i g_{a\oo b} - A_{ia}{}^cg_{c\oo b} - g_{a\oo c}A_{i\oo b}{}^{\oo c}
= \del_i g_{a\oo b} - A_{ia}{}^cg_{c\oo b}$$ (the mixed index parts of the
connection vanish by the K\"ahler condition). Thus,
\eqn\econn{A_i = \del_i g\cdot g^{-1}
= - g\del_i g^{-1}.}
The $tt^*$ equations are derived by path integral
manipulations like the ones used in finding the holomorphic basis.  In
fact, the existence of such a basis immediately tells us that the chiral
ring matrices $C_j$ (defined by $\phi_i\phi_j = (C_i)_j{}^k\phi_k$) obey
$D_{\oo i}C_j = \del_{\oo i}C_j = 0,$ since the chiral ring has no $\oo
t_{\oo i}$ dependence -- the $\oo t$ terms in the action are $Q^+$-trivial.
Similarly one shows that the topological metric, $\eta,$ only has
holomorphic dependence on the couplings.

We can succinctly express the $tt^*$ equations by considering a familry of
connections indexed by a ``spectral parameter,'' $x:$
\eqn\conns{\eqalign{\nabla_i &= D_i - xC_i,\cr
\overline{\nabla}_{\overline{i}} &= \overline{D}_{\overline{i}} -
x^{-1}\overline{C}_{\overline{i}}.}}
The $tt^*$ equations, conditions on the metric and the $C_i,$ are then
summarized by the statement that
$\nabla$ and $\overline{\nabla}$ are flat for all $x.$ For example, from the
term mulitplying $x$ in $[\nabla_j,\oo{\nabla}_{\oo i}] = 0$ we have the
formula proved above: $$[D_{\oo i},C_j] = D_{\oo i}C_j = 0.$$  The terms
independent of $x$ in the same equation give
$$[D_j,D_{\oo i}] + [C_j,\oo C_{\oo i}] = 0.$$
In the holomorphic basis, we know what the connection is from \econn.  Further,
one finds the action of $\oo C_{\oo i}$ is by the matrix $gC_i^\dagger g^{-1}.$
 Thus, we arrive at
$$\overline{\del}_{\overline{i}}(g\del_j g^{-1}) =
[C_j,gC_i^\dagger g^{-1}].$$

In the foregoing, we have assumed that the two-dimensional space was an
infinite cylinder with unit perimeter.  A perimeter of $\beta,$ which we
will take as a scaling parameter, adds a factor of $\beta^2$ to the right
hand side.  Now flow in the space of theories along the $\beta$ direction -- a
change in scale, that is -- is given by the renormalization group.  We can
study what the connection
in the $\beta$ direction looks like.  One finds that the gauge field in
the $\beta$ direction corresponds to the Ramond charge matrix, $q,$ in
the conformal limit \rNEWSUSY.  We define
\eqn\Qmat{Q_{ab} = -\half (g\beta\del_\beta g^{-1})_{ab}}
(we have actually taken the direction defined
by $\tau,$ where $\beta = \e{(\tau + \tau^*)/2}$).  Then $q = Q$ as $\beta
\rightarrow 0.$
The
reason for this relation is that under a scale transformation, a state --
represented by a path integral with a circle boundary -- changes by the
trace of the energy momentum tensor plus the integral along the boundary of the
topological twisting background gauge field coupled to the chiral
fermion current:
$$\delta g = \epsilon g \Rightarrow \delta\vert a\rangle = \vert\int{\rm Tr}
T^{\rm (top)}\phi_a\rangle = \vert\int T_\mu^\mu \epsilon\phi_a\rangle +
\vert\int\partial_\mu J^{\mu}\phi_a\rangle.$$
The matrix $Q$ is the axial (left plus right) charge
matrix, which is a conserved charge only at the conformal point (for example,
any mass in the theory breaks this chiral $U(1)$).
If we change our point of view, exchanging ``space''
and ``time'' in the path integral expression for $Q,$ then the configurations
being
integrated are the solitons which run from vacuum $a$ to vacuum $b$ along the
spatial line.  The
chiral fermion number then gets rewritten as the fermion number, since
$j^5_\mu = i\epsilon_{\mu\nu}j^\nu$ in two dimensions.  The result is
expressed as a limit as the spatial volume goes to infinity:
\eqn\newind{Q_{ab} = \lim_{L\rightarrow\infty}i{\beta\over
L}\Tr_{ab}(-1)^FF\e{-\beta
H}.}
The $ab$ subscript indicates that the trace should be performed over the
$ab$ soliton sector of the Hilbert space.  It is clear from the presentation
\newind\ that the coefficient of the leading exponential for large $\beta$ can
be used to count the minimum energy solitons between $a$ and $b,$ weighted by
$F(-1)^F.$

Let us now consider the following set of equations: \eqn\lax{\nabla_i
\Psi(x,w_a) = \overline{\nabla}_{\overline{i}}\Psi(x,w_a) = 0.} In order
to solve these equations simultaneously, we must require that $\nabla$
and $\overline{\nabla}$ commute, i.e. they are {\sl flat}; the
consistency condition is thus $tt^*$.  We can amend the connection to
include the variable $x$ so that the independence of the phase of $\beta$
follows from requiring flatness.
That this independence should hold
follows from the freedom to redefine the phases of the fermions,
eliminating an overall scale in the superpotential.\footnote{$^*$}{By
``superpotential,'' we mean the values $w_a$ which can be assigned to the
different vacua such that the Bogolmonyi soliton masses (the central
terms in the N=2 algebra on the non-compact line) are given by the absolute
values of
differences of the $w_a$ (these are the canonical coordinates of the
theory).  Note that $\beta = \e{-A},$ where $A$ is the one-instanton
action or area.}
We write:
\eqn\amendtt{x{\del\over\del
x}\Psi = (\beta xC + Q - \beta x^{-1}\oo C)\Psi.}
The matrix $C$ is given by $C_i{}^j = \sum_{k}w_kC_{ki}{}^j$
(see footnote for definition of $w_k$)
$\oo C = gC^\dagger g^{-1},$ and $Q$ is as above.
Here we have written
that the scale of the superpotential is $\beta\e{i\theta}$ with $x =
\e{i\theta},$ but the equations now make sense for $x$ a complex
variable.  In general, there will be $n$ solutions to \lax, so we take
$\Psi$ to be an $n\times n$ matrix whose columns are solutions.  The
equations are singular at $x=0,\infty,$ which means the columns of $\Psi$
will mix under monodromy $x \rightarrow \e{2\pi i}x$:  $\Psi \rightarrow
H\cdot\Psi.$  In fact, the solutions can be expressed in terms of two
regions of the $x$-plane.  In the overlap of the regions, the solutions
are matched by matrices which are related such that the total
monodromy takes the form $$H = S\cdot (S^{-1})^t.$$ Furthermore, a flat
connection guarantees that $H$ (as well as $S$) is a {\sl constant}
matrix -- independent of local variations the couplings of the theory.
In particular, we can
evaluate it in a convenient limit.

Consider $\beta \rightarrow 0$ with $x$ small; this is the conformal
limit, since the area goes to infinity.  (We note that the scale of $\beta,$
since $\ln{\beta}$ multiplies the actions, does not affect the configuration of
the vacua.)  In this limit, the equation in $x$
takes the simple form $${d\over d\theta}\Psi_i = q_{ij}\Psi_j,$$ where
$\theta$ is the phase of $x$ and $q = Q\vert_{\beta=0}$ is the Ramond charge
matrix. The solution is $\Psi(\theta) = \e{2\pi i q
\theta}.$  It is then clear that $${\rm Eigenvalues}[H] = {\rm
Eigenvalues}[\e{2\pi i q}]:$$ the phases of the eigenvalues of the
monodromy around zero are precisely the Ramond charges.  Since these
charges must be real, the eigenvalues $\lambda_i = \e{2\pi iq_i}$ of the
monodromy must satisfy $\vert \lambda_i \vert = 1.$

We can now go to the infrared limit ($\beta$ large) to determine $H,$ as it is
independent of $\beta.$  Let us see how the (weighted) numbers of solitons
enter in the calculation of $H.$  As we discussed when deriving \newind, these
numbers appear as terms in the leading asymptotics of the matrix $Q_{ab}.$  In
the large $\beta$ limit, the soliton states of minimal energy -- the Bogolmonyi
solitons -- dominate the expression for $Q,$ so that the leading behavior is
$$Q_{ij}\big\vert_{\beta\rightarrow\infty} = -{i\over
2\pi}A_{ij}m_{ij}\beta K_1(m_{ij}\beta),$$ where $m_{ij}$ is the soliton
mass. Using the relation \Qmat\ between $Q$ and $g^{-1}\del g,$ the asymptotic
form of $g$ is seen to be
\eqn\solnum{g_{i\oo j} = \delta_{ij} - {i\over
\pi}A_{ij}K_0(m_{ij}\beta),}
where the $K_0$ and $K_1$ are modified
Bessel functions.  Soliton numbers thus are directly related with solutions to
the $tt^*$ equations.  The more rigorous analysis of section 4 of \rCVNEW is
needed to relate the solutions $\Psi$ of \lax\ to the metric $g_{i\oo j}.$  The
monodromy $H$ is found to be related to the matrix $A$ by the following
expression:
\eqn\monod{\eqalign{H &= S(S^{-1})^t, \cr S &= 1 - A.}}
In a standard configuration of vacua, $S$ is an upper-triangular matrix.

We have therefore seen that the soliton numbers counted with $F(-1)^F$ can be
arranged in a monodromy matrix $H = (1-A)(1-A)^{-t}$ whose eigenvalues give the
chiral charges of
the vacua in the conformal limit (the integer part of the phases are
defined by smoothly varying the identity matrix to $A$ while counting the
number of times the eigenvalue winds around the origin).  We get
constraints on $H$ due of these facts.  It must be integer valued.  Its
eigenvalues ${\lambda_i}$ must obey $\vert\lambda_i\vert = 1.$  Their
phases must lie symmetrically around zero, due to fermion number conjugation.
In addition, there is an
action of the braid group, corresponding to changes in couplings which
alter the configuration of the vacua in the $W$ plane (defined abstractly
for non-Landau-Ginzburg theories -- see previous footnote) and hence the number
of solitons connecting them.
Specifically, the Diophantine constraints are that the characteristic
polynomial of the $n\times n$ matrix $H,$
$$P(z) = {\rm det}(z - H)$$ must obey
$$P(z) = \prod_{m\in {\bf N}}\left( \Phi_m(z)\right)^{\nu(m)},$$
where $\nu(m)\in {\bf N}$ (non-negative integers) are almost all zero, and
$\Phi_m(z)$ is the cyclotomic polynomial of degree equal to $\phi(m),$ the
number of numbers relatively prime to $m.$  Further \rCVNEW,
$$\eqalign{&1.\> \sum_{m}\nu(m)\phi(m) = n\cr
&2.\>\nu(1) \equiv 1\, {\rm mod}\, 2\cr
&3.\>n \in 2{\bf Z} \Rightarrow \> either \> \nu(1)>0 \> or\>
\sum_{k\geq1}\nu(p^k)\equiv 0\, {\rm mod}\, 2, \> \hbox{for all primes}\> p.}$$

We are primarily interested in the sigma model case, for which the Ramond
charges lie in the set $\{-d/2, -d/2 + 1,....,d/2 -1, d/2\},$ where $d$ is the
dimension of the K\"ahler manifold, $M.$  Further, we restrict our attention to
manifolds with diagonal Hodge numbers (or else the finite chiral ring would
have nilpotent elements of non-zero fermion number, and no canonical basis --
crucial to the derivations -- would exist).  In other words,
$$P(z) = (z-\epsilon)^{\chi(M)},$$
where $\epsilon = 1$ for $d$ even and $\epsilon = -1$ for $d$ odd.

Let us now illustrate some solutions to these equations.  The first example has
an obvious physical interpretation; we will return to discuss the next example
-- affine Lie groups -- in section 2.4.  The simply laced Lie groups are
related to possible solutions for $A$ as
follows.  Suppose the matrix $B = S + S^t$ is positive definite.  Then
$HBH^t = SS^{-t}(S + S^t)S^{-1}S^t = B,$ which means that $H$ is in the
orthogonal group to the quadratic form, $B,$ which tells us that $H$ is
simple and $\vert \lambda_i \vert = 1.$  The simply laced Lie groups
correspond to positive definite integral matrices through their Cartan
matrices.
$B$ defines an inner product
on ${\bf R}^n,$ and if we take $A$ to be upper triangular, with
$A_{ij} = -B_{ij}/2, i < j,$ then $H = (1-A)(1-A)^{-t}$ satisfies the
Diophantine
constraints.  These matrices correspond to the N=2 {\sl A-D-E} minimal models,
and have explicit realizations as Landau-Ginzburg theories.
Weyl reflections of the lattice vectors produce different, though
equivalent solutions to the Diophantine equations.  These reflections
correspond to perturbations of the superpotential, $W,$ such that the vacua
move through colinear configurations
in the $W$ plane.  Such reconfigurations of the vacua produce a braid group
action on the matrix $H.$

The affine Lie groups correspond to the case where $B = S
+ S^t$ has a single zero eigenvector, $v,$ thus satisfying $S^tv = -Sv.$
Then $B$ defines a reduced
matrix $\hat{B},$ on the orthogonal complement to ${\bf R}v,$ which solves
the Diophantine equations.
We now note that
$$H^t v = S^{-1}S^{t}v = -S^{-1}Sv = -v,$$
so $\vert\lambda_v\vert = 1$ and we
see that all the
eigenvalues $\lambda$ of $H$ indeed have $\vert \lambda \vert = 1.$  The
remaining constraints on $H$ are satisfied as well.

\subsec{Mutations}
We have already discussed a monodromy when one of the parameters of the theory
is taken around the origin.  In fact there are a number of discrete mutations
or braidings of these theories.  The treatment in this paper has been for
general N=2 theories, but this braiding is particularly intuitive in the case
of Landau-Ginzburg models (all the results are valid in the general case).
These models are described by a superpotential $W(X)$ and the vacua correspond
to critical points $X_i$ such that $\nabla W(X_i) = 0.$  The locations of these
points clearly depends on the parameters of $W$.  Solitons, it is found, travel
on straight lines in the $W$ plane, and so a discrete shift in soliton number
can occur when these vacua pass through a colinear configuration.  This
situation describes precisely the Picard-Lefschetz theory of vanishing cycles,
(the inverse image of a point along the line in the $W$ plane is a homology
cycle, and solitons correspond to intersections of two homology
cycles)\footnote{$^*$}{We note here that the mathematical theory to be
discussed has been seen to parallel the Picard Lefschetz theory as well, though
a greater understanding of the relation is still unknown.} which undergo basis
changes when crossed.  Essentially, the change is $$A_{ac} \rightarrow A_{ac}
\pm A_{ab}A_{bc} \> \hbox{(no sum)},$$ where the sign depends on the
positive/negative orientation of the crossing.  In a configuration in which the
matrix $S$ is upper triangular, the change of basis matrix implies
$$S \rightarrow PSP,$$
where $P = \pmatrix{0&1\cr 1&-S_{ij}}$ in the $ij$ subsector.  Note that $P$
depends on $S$ itself, and so the mutation is nonlinear.

We note here, too, that the canonical basis is only defined up to a sign.
Further, reversing the orientation of the $W$ plane -- equivalently, taking the
monodromy in the other direction -- leads to $H \rightarrow H^{-1},$ i.e.
$$S\rightarrow S^t.$$
Thus, all matrices $S$ obtained by any combination of the above transformations
are related to the same (continuum class of theories associated to the) N=2
quantum field theory.

Finally, we show how $S$ and $S^{-1}$ can be related.  To the braid group of
$n$ objects, generated by $P_i, i = 1...n,$ which denote braids of the $i^{\rm
th}$ object over the $(i+1)$-th object, we define the element
$$\nu = P_1P_2P_1P_3P_2P_1...P_{n-1}P_{n-2}...P_2P_1$$
consisting of $\pmatrix{n\cr 2}$ transformations.  This corresponds to
reversing the orders of the elements.  (Note that as a matrix $\nu$ depends
nonlinearly and nontrivially on $S.$)  Then if $J = \delta_{i,n+1-i}$ is a
reordering, we find
$$S^{-t} = J\nu S\nu J.$$
In particular, $S$ and $S^{-1}$ are associated to the same N=2 theory.  We will
use this point in our comparisons of results from math and physics.

We conclude this subsection with a simple exercise.  We take $S =
\pmatrix{1&a&b\cr 0&1&c\cr 0&0&1}.$  Then $P_1$ is represented for this $S$ by
$P_1 = \pmatrix{0&1&0\cr 1&-a&0\cr 0&0&1}.$  One finds from $P_1SP_1$ that the
action is $P_1: (a,b,c)\mapsto (-a,c,b-ac).$  Similarly, $P_2: (a,b,c)\mapsto
(b,a-bc,-c).$  The element $\nu = P_1P_2P_1$ then sends
$$(a,b,c)\mapsto (-a,c,b-ac)\mapsto (c,-a-c(b-ac),ac-b)\mapsto (-c,ac-b,-a).$$
Therefore, $\nu: S\rightarrow \widetilde{S} = \pmatrix{1&-c&ac-b\cr 0&1&-a\cr
0&0&1},$ and it is simple to check that $J\widetilde{S}J = S^{-t}.$  Thus $S$
and $S^{-1}$ are related (taking the transpose, as discussed above).

Of course, for the general N=2 theory there is no simple geometric
interpretation of the placement of the vacua, though colinearity is still
well-defined in terms of the N=2 algebra.  Still, the question of how
perturbations of the theory affect the vacua is quite subtle.  In addition,
some of the perturbations may not make sense physically.  For example,
perturbations by nonrenormalizable terms are not allowed.  We suspect that
these phenomena are related to the question of constructability of helices, an
issue we will return to from the mathematical viewpoint in section 3.

\subsec{Examples:  Projective Spaces, Grassmannians, and Orbifolds}
One of the few cases in which we can compute the soliton matrix by studying
$tt^*$ asymptotics directly is the simple case of the projective line ${\bf
P}^1.$  This theory has two chiral ring elements corresponding to its
cohomology.  Let us label them $1$ and $X.$  The quantum ring is $X^2 = \beta
\equiv {\rm e}^{-A},$ where $A$ is the area of the ${\bf P}^1,$ complexified so
as to include the $\theta$ angle.  Briefly, this comes about as follows.  $X$
has a non-vanishing one-point topological correlation (which we normalize to
one), since there is a unique constant map taking the insertion point on the
sphere to the chosen point on ${\bf P}^1$ representing $X,$ the volume form.
The only non-vanishing three-point correlator is $\langle XXX\rangle = \beta,$
which has nonzero contribution at instanton number one arising from the unique
holomorphic map from the (Riemann surface) sphere to the (target space) sphere
taking the three insertion points to three specified points.  This gives $X^2 =
\beta.$

The metric $g_{a\oo{b}}$ is diagonal, due to a ${\bf Z}_2$ symmetry which is
the leftover of the anomalous $U(1)$ symmetry (the chiral $U(1)$ is broken to
${\bf Z}_n$ on ${\bf P}^{n-1}$).  This tells us that $\langle\oo{1}\vert
X\rangle = \langle\oo{X}\vert 1\rangle = 0.$  Thus $g = \pmatrix{a&0\cr 0&b},$
with $a$ and $b$ real, as $g$ is hermitian.  The metric $\eta_{ab}$ is given by
$\eta = \pmatrix{0&1\cr 1&0},$ since $X$ only has a non-vanishing one-point
correlation.  The reality constraint, or $CPT,$ tells us
$$\eta^{-1}g(\eta^{-1}g)^* = \pmatrix{0&1\cr 1&0}\pmatrix{a&0\cr 0&b}
\pmatrix{0&1\cr 1&0}\pmatrix{a&0\cr 0&b} = {\bf 1},$$
which gives $b = a^{-1}.$

We write the $tt^*$ equations for variations with respect to $\beta$ and
$\oo{\beta}.$  We thus need the matrix $C_{\beta},$ corresponding to the
operator represented by varying $\beta.$  Clearly $A$ is the coefficient for
$X,$ and so $C_\beta = C_X\left({\partial A\over\partial\beta}\right) =
-{1\over\beta}\pmatrix{0&1\cr \beta&0};$ also, $gC_{\beta}^{\dagger}g^{-1} =
\pmatrix{0&\oo{\beta}a^2\cr a^{-2}&0}.$  The $tt^*$ equation is then
\eqn\ponett{\del_{\oo{\beta}}\pmatrix{a&0\cr
0&a^{-1}}\del_{\beta}\pmatrix{a^{-1}&0\cr 0&a} = {1\over
\vert\beta\vert^2}\left[\pmatrix{0&1\cr \beta&0},\pmatrix{0&\oo{\beta}a^2\cr
a^{-2}&0}\right].}
Both nontrivial components of \ponett\ are equivalent.  Further, $a$ only
depends on the absolute value of $\beta,$ since the phase can be absorbed by a
redefinition of the phase of the fermions \rCVTAF.  Let $x=\vert\beta\vert$ and
define $u=\ln{(a^2x)}.$
Then \ponett\ gives
$$u'' + {1\over x}u' = 4{\rm sinh} u.$$

As often happens, requiring finiteness of the metric at $x=0$ fixes the metric
(i.e. one boundary condition is enough to impose).  We need
$$u\rightarrow\ln{x} + c$$ as $x\rightarrow 0.$  The asymptotic behavior at
$x\rightarrow\infty$ has been analyzed in \rPIII.  We compute the soliton
number from the asymptotics of $g$ in the canonical basis $O_{\pm} =
(X\pm\sqrt{\beta}).$  Extracting the lone soliton number $A_{\pm}$ from
\solnum, we find
$$S = \pmatrix{1&-2\cr 0&1}; \qquad H = SS^{-t} = \pmatrix{-3&-2\cr 2&1}.$$
Note that ${\rm det}(z-H) = (z+1)^2 = \Psi_2(z)^2$ ($\Psi_2$ denotes the second
cyclotomic polynomial), which gives the Ramond charges $N + \half.$  The
integer part, $N,$ can be determined by recording how many times the phases of
the eigenvalues of $H(t)$ wrap around the origin as $S(t) = \pmatrix{1&-2t\cr
0&1}$ runs from the identity matrix to $S$ while $t$ spans the interval.  One
easily calculates that the two phases are $$\theta_{\pm} = \pm{\rm
tan}^{-1}{2t\sqrt{1-t^2}\over 1-2t^2},$$
and so the charges are $\pm\half$ as they should be.  The braid group action is
simple in this case.  There is one mutation, $P_1,$ which sends the matrix
element $S_{12} \rightarrow -S_{12}$ (this can also be effected by a change of
sign).

The classification program, in all its glory, has been illustrated by this
simple example.  Other spaces, such as the higher projective spaces and
Grassmannians, are too unwieldy for a direct analysis.  Too little is known
about the solutions to the $tt^*$ equations, which for ${\bf P}^n$ correspond
to affine Toda equations.  Perhaps the proposed relation between math and
physics is best borne out by a rigorous analysis of these equations and their
asymptotic properties.

Fortunately, the soliton numbers for ${\bf P}^{n-1}$ are computable by other
methods, and the Grassmannians $G(k,N)$ ($k$-planes in ${\bf C}^N$) may be
analyzed as well.  First let us consider ${\bf P}^{n-1}.$  These theories have
an anomalous $U(1)$ charge (which is evident in the chiral ring, which has the
simple form $X^n = \beta$).  Instantons break this symmetry down to ${\bf
Z}_n.$  We can choose a basis for the vacua such that the ${\bf Z}_n$ symmetry
cyclically rotates the $n$ vacua.  The soliton matrix $\mu_{ab}$ then depends
only on the difference $b-a.$  Further, one can directly analyze the properties
of the Stokes matrix by studying possible asymptotic solutions to \lax\ and
\amendtt \rCVNEW.  These considerations allow us to write the monodromy matrix
$H$ as $H = AB,$ where $A$ commutes with $H.$  This condition allows us to
conclude that the characteristic polynomial of $B$ must also contain only
products of cyclotomic polynomials.  The real information, however, comes in
the fact that $B$ encodes the soliton numbers as well.

To see how this works, let us simply note that the Ramond charges of ${\bf
P}^{n-1}$ have the form $q_k = k - (n-1)/2$ and are thus half-integral or
integral when $n$ is even or odd, respectively.  We therefore have
$${\rm det}(z-H) = (z\pm 1)^n$$
for $n$ even/odd.
As a result, we conclude that, for $n$ odd, say,
${\rm det}(z - B) = (z-1)^n = \sum_{k = 0}^{n}\pmatrix{n\cr k}(-1)^kz^k,$
from which we may conclude
\eqn\pnnums{S_{ij}= (-1)^{j-i}\pmatrix{n\cr j-i}.}
The minus signs may be removed by a redefinition of vacua ${\rm e}_a
\rightarrow (-1)^a{\rm e}_a,$ but we shall leave them in.  Similarly, for $n$
even we get the same result without the minus signs.  We can reinsert them by
performing the same change of basis.

These models have the special property of being integrable, for a special
choice of the K\"ahler metric; i.e., they have an infinite number of conserved
quantities such that the momenta of solitons are only permuted by interactions.
 Interactions of solitons in integrable models can be computed using the
thermodynamic Bethe ansatz.  Such an analysis shows the lightest solitons
appear in fundamental multiplets of the original $SU(n)$ symmetry.  These
represent the $n$ solitons of $\mu_{i,i+1}.$  The interpretation of the other
solitons of greater fermion number is that they are particles formed by
anti-symmetric combinations of these $n$ solitons.\rKen  Witten considered
these solitons in Ref. \rWCPN, in which he formulated the ${\bf P}^{n-1}$ model
by gauging a $U(1)$ action on fields $n^i \in {\bf C}^n$ constrained by
Lagrange multipliers to satisfy $\vert n\vert^2 = 1.$  The effective theory of
the Lagrange multipliers and gauge field relates the topological charge of
solitons to the $U(1)$ charge of the gauge field via the Gauss' law equation of
motion.  These $n$ fields $n^i$ are the fundamental solitons connecting
neighboring vacua, and clearly transform under the fundamental representation
of $SU(n).$  Other solitons are (anti-symmetric) composities of these fields.

The Grassmannian case is more subtle, and we must use a different technique,
discussed in section 8.2 and the appendix A of  Ref. \rCVNEW.  The Grassmannian
$G(k,N)$ of $k$-planes in ${\bf C}^N$ is a complex manifold of dimension
$k(N-k),$and can be identified with the homogeneous space $U(N)/(U(k)\times
U(N-k)).$  Cecotti and Vafa have shown how to relate the observables to $k$
copies of ${\bf P}^{N-1}.$  The prescription is to take as vacua fully
anti-symmetric tensor products of $k$ vacua for ${\bf P}^{N-1}.$  There are
$\pmatrix{N\cr k}$ such choices, equal in number to the Euler class of
$G(k,N).$  The inner product of two vacua is then given in terms of the
constituent ${\bf P}^{N-1}$ vacua.  In this manner, the Grassmannian case can
be reduced to the projective spaces.  However, in the Grassmannian case, there
is an ambiguity in asking what the soliton numbers are, as various vacua are
aligned in the $W$ plane.  Such a configuration can lead to non-integer entries
in the matrix $S$ yielding monodromy matrices $H$ which do not satisfy the
classification equations.  In the case of the Grassmannians, the matrix
so-obtained does not satisfy the Diophantine equations, presumably for this
very reason.  To fully resolve this difficulty, one would need to perturb the
model so that the vacua were configured with no three of them colinear.

Orbifolds of ${\bf P}^1 \cong S^2$ by discrete subgroups of $SO(3)$ are
interesting cases in which the $tt^*$ equations may be explicitly used to
compute soliton numbers.  This procedure was employed in \rCVMO\ and \rZDD.
What makes these theories workable is that the orbifold theory possesses a
symmetry which simplifies the metric, $g_{a\oo{b}}.$  The only new subtlety is
in the computation of the quantum rings, which are obtained by an analysis of
equivariant holomorphic maps from covering surfaces, branched over the
insertion points of twisted observables.  This theory was developed in \rHV\
and \rZAZ.\footnote{$^*$}{The orbifold of a quantum field theory -- a model
with target space $M/G,$ where $G$ is a discrete group acting on $M,$ includes
states in the Hilbert space which correspond to strings running between points
on $M$ related by an element of $G.$  The holomorphic maps, or instantons, to
this singular space (as $G$ may have fixed points) are analyzed by studying
holomorphic maps between $G$-covers, which are equivariant with respect to the
$G$ action.}

The case of the dihedral groups yields the following results, easily
generalizable though not yet proved for the tetrahedral, dodecahedral, and
icosahedral groups \rZAZ.  A discrete subgroup $G$ of $SO(3)$ lifts under the
${\bf Z}_2$ covering $SU(2)\rightarrow SO(3)$ to a subgroup $\widetilde{G}$ of
$SU(2),$ which is associated to a Dynkin diagram in the following way, due to
McKay \rSLOD.  The fundamental representation of $SU(2)$ defines a
two-dimensional representation $R$ of any subgroup.  If we label the
irreducible representations of $widetilde{G}$ by $V_i,$ we can define a matrix
$A_{ij}$ by the tensor decomposition
$$V_i\otimes R \cong \oplus_{j}N_{ij}V_j.$$
McKay's theorem states that the matrix $N$ is the adjacency matrix of a Dynkin
diagram of an affine Lie algebra.  As $N$ is symmetric with zeroes on the
diagonal, we can write $N = A + A^t,$ where $A$ is upper triangular.  For
example, to compute the matrix for the ${\bf Z}_N$ orbifold of a sphere, we
compute the matrix $N$ for the double cover ${\bf Z}_{2N}.$  The Dynkin diagram
corresponds to the Lie group $\widetilde{A}_{2N-1},$ and looks like a circular
chain of $2N$ dots. For the dihedral groups $D_N,$ the matrix $A$ obtained from
the this procedure yields the Dynkin diagram for the affine Lie group
$\widetilde{D}_{N+2}$ (we should not be disconcerted by the mismatch of
numbers; there need be no relation).  This analysis is verified physically by
computing asymptotics of $g_{a\oo{b}}$ from the $tt^*$ equations.

Of course, we have already proven that these affine Dynkin diagrams yield
solutions to the classification constraints.   The discussion here allows us to
identify these solutions as orbifolds.

\newsec{The Math}
\subsec{Overview}

We have stressed that no link has been found between the physics we just
discussed and the mathematics we will introduce, so our motivation is indirect.
 Nevertheless, the evidence that some link exists is compelling.  Certainly,
none will be found without a thorough understanding of the structures at hand.

The evidence is the following.  In the previous section we constructed, given a
K\"ahler manifold with positive first Chern class (to guarantee asymptotic
freedom, so that the quantum field theory makes sense) and diagonal Hodge
numbers (so that a canonical basis exists), a quasi-unipotent matrix satisfying
certain Diophantine equations regarding its eigenvalues.  This matrix -- the
matrix of (properly counted) soliton numbers between vacua -- had an action of
the braid group on it.  The new matrices so constructed also satisfied the
equations in question.  Here we will start with a toplogical space and consider
sheaves over that space (for the time being we can think of the sheaves as
vector bundles).  We will construct a set of ``basis'' sheaves, i.e. a set such
that all other sheaves are ``equivalent'' to (i.e. have a resolution in terms
of) a direct sum of sheaves from the basis set.  Now from this set we simply
consider the bilinear form which is the Euler character between two sheaves,
i.e. the alternating sum of dimensions of cohomology classes.  This matrix will
be quasi-unipotent.  The choice of basis set will not be unique, and the
different choices will yield different matrices which satisfy the same
properties.  Further, we will be able to show for the projective spaces that
the matrices are {\sl exactly the same} as in the physics case.  We will also
explore other examples which have not yet been solved physically.

To make this clear, let us learn about coherent sheaves, helices, and braiding.
 Finally, we will look at some examples -- projective spaces and Grassmannians
-- in detail.

\subsec{Coherent Sheaves}
We begin with a brief discussion of sheaves, following a brief summary of the
algebraic geometry we will need.  The treatment here borrows liberally from
Ref. \rGH. Heuristically, we can think
of a sheaf as the generalization of a vector bundle when we replace a
vector space by an abelian group.  Thus, given a topological space, $X,$
a sheaf $\sF$ on $X$ gives a set of sections $\sF (U)$ which are abelian
groups with the following properties.  Given open sets $U\subset V$ we
have a restriction map $r_{V,U}:\sF (V)\rightarrow\sF (U)$ satisfying for all
$U\subset V\subset W,$ $\sigma\in\sF (A), \> \tau\in\sF(B),$
$$\eqalign{&\bullet \> r_{V,U}\circ r_{W,V} = r_{W,U} \qquad(\hbox{thus we can
write } \>\sigma\vert_U\>
\hbox{for}\> r_{V,U}\sigma )\cr
&\bullet\> \sigma\vert_{A\cap B} = \tau\vert_{A\cap
B}\Rightarrow\>\exists\,\rho\in\sF(A\cup B)\>{\rm
s.t.}\>\rho\vert_{A}=\sigma,\>\rho\vert_{B}=\tau\cr}$$
Roughly speaking, this says that sections are determined by their
values on open sets.

The most important example of a sheaf for us will be $\O_n,$ the sheaf of
holomorphic functions in $n$ variables (usually, we will drop the subscript
$n$).  Thus, $\O(U)$ consists of the ring of holomorphic functions on $U\subset
{\bf C}^n.$  Clearly $\O(X),$ when $X$ is a compact complex manifold, is equal
to the globally holomorphic functions, or constants.

The algebra of sheaves will be important to us, so we briefly review the
pertinent aspects.  We should have little difficulty with the
constructions, as they closely parallel manipulations of abelian
groups or vector bundles.
A map between sheaves $$f:\sF\rightarrow\sG$$ over $X$ is a
collection of group homomorphisms\footnote{$^*$}{Recall a homomorphism $f$
between abelian groups $A, B$ is a map commuting with the group addition:
$f(a+b)=f(a)+f(b).$} $f_U:\sF(U)\rightarrow\sG(U).$  As with abelian
groups, given such a sheaf map we can define its kernel and cokernel.
The sections of the kernel are simply the kernels of the
section maps $f_U.$  Thus, $\ker(\sF)(U) = \ker(f_U).$  The cokernel is
slightly subtle.  We cannot take the na\"\i ve definition of $\cok,$ which
would read $\cok(\sF)(U) \hbox{ ``='' } \cok(f_U),$ as it does not satisfy the
properties required of sheaves.  For example, consider the sheaves $\cal O$ and
$\cal O^*$ of holomorphic and non-vanishing holomorphic maps on ${\bf C}-0.$
The abelian groups are addition and multiplication, respectively.  Then the
exponential map $\exp: h(z) \mapsto \e{2\pi ih(z)}$ is a homomorphism, with
kernel ${\bf Z},$ the sheaf with integer-valued sections.  Generalizing the
notions of algebra, we would like to have the sequence
\eqn\seq{0\longrightarrow {\bf Z}\matrix{ {}_{i}\cr\longrightarrow \cr {} } \O
\matrix{ {}_{\rm exp}\cr\longrightarrow \cr {} }\O^*\longrightarrow 0}
be exact, i.e. have $\O^* = \O/{\bf Z} = \cok(i).$
However, with our simple definition, the cokernel is not even a sheaf.
Consider the section $h(z) = z \in {\cal O}^*({\bf C}-0)$ which is not in the
image $\exp(\O({\bf C}-0)).$  Clearly, its restrictions to the contractible
sets $U_1 \equiv \{ -\epsilon < {\rm Arg}\, z < \pi +\epsilon\}$ and $U_2
\equiv \{ \pi - \epsilon < {\rm Arg}\, z < \pi\}$ are in the image of $\exp :
{\cal O}(U_i)\rightarrow {\cal O}^*(U_i), \> i=1,2.$  Thus the second condition
above is not satisfied.
To remedy this, we define the sections of the cokernel of $f $ over $U$ to be
given by a set of sections $\sigma_\alpha$ of a cover $\{U_\alpha\}$ of $U$
satisfying
$$\sigma_\alpha\vert_{U_\alpha\cap U_\beta} - \sigma_\beta\vert_{U_\alpha\cap
U_\beta} \in f(\sF(U_\alpha\cap U_\beta)) .$$
We then identify two sections (i.e. collections $(U_\alpha,\sigma_\alpha)$ and
$(V_\beta,\rho_\beta)$) if at all points $p$ in $U$ and open sets $U_\alpha,$
$V_\beta$ containing $p,$ there exists a neighborhood $N\subset U_\alpha\cap
V_\beta$ such that the restrictions to this neighborhood of $\sigma_\alpha$ and
$\rho_\beta$ differ by the image of a section on $N.$  This procedure allows
the cokernel to be defined on open sets which would otherwise be ``too big,''
by allowing the equivalence to be true up to refinements.
It can be checked that the sequence \seq\ is now exact for any complex
manifold, i.e. $\O^* = \cok(i).$

Generally, a sequence of sheaves
$$...\sF_{n-1}\matrix{ {}_{f_{n-1}}\cr\longrightarrow\cr{}}
\sF_n\matrix{ {}_{f_n}\cr\longrightarrow\cr{}}\sF_{n+1}\longrightarrow...$$ is
said to be exact if $f_{n}\circ f_{n-1} =0$ and
$$0\longrightarrow\ker(f_{n-1})
\longrightarrow\sF_{n-1}
\longrightarrow\ker(f_n)
\longrightarrow 0$$
is exact for all $n.$  We will henceforth assume that a fine enough cover
$(U_\alpha)$ exists such that the subtleties discussed are erased, i.e. such
that each induced sequence of sections over $U_\alpha$ is exact.

We are interested in topological properties of the sheaves we will consider.
Sheaf cohomology measures global properties of sheaves by comparisons on
intersections of a cover.  Let $(U_\alpha)$ be a cover of a manifold, $M,$ and
$\sF$ a sheaf over $M.$   Then we define the sheaf cohomology by taking the
cohomology of the following complex.
Define $C^m$ to be the disjoint union of the sections of all $(m+1)$-fold
intersections of the $U_\alpha:$
$$C^m(U,\sF) = \prod_{\alpha_i\neq\alpha_j}\sF(U_{\alpha_0}\cap...\cap
U_{\alpha_m}).$$  Then the coboundary $\delta: C^m(U,\sF)\rightarrow
C^{m+1}(U,\sF)$ is defined by
$$(\delta\sigma)_{\alpha_0,...,\alpha_{m+1}} = \sum_{j=0}^{m+1}(-1)^j
\sigma_{\alpha_0,...,\hat{\alpha_j},...,\alpha_{m+1}}
\vert_{U_{\alpha_0}\cap...\cap U_{\alpha_{m+1}}}.$$
The sheaf cohomology $H^*(M,\sF)$ is just the cohomology $\ker(\delta)/{\rm
Im}(\delta)$ of this complex, provided we choose an appropriately fine
cover.\footnote{$^*$}{The actual sheaf cohomology is defined as a limit of the
stated cohomologies under refinements of the cover.  If the cover is {\sl
acyclic,} meaning the cohomology of any multiple intersection is trivial (for
example if they are contractible for sheaves of holomorphic $p$-forms), then
the sequence yields the proper sheaf cohomology.}  As an example, we observe
that $H^0(M,\sF)= \ker(\delta)\subset C^0,$ which is the set of collections
$\{\sigma_\alpha\}$ obeying $(\delta\sigma)_{\alpha\beta} = \sigma_\beta -
\sigma_\alpha = 0$ (with the restriction to $U_\alpha\cap U_\beta$ understood).
 This is precisely the data which determines a global section.  Thus,
$H^0(M,\sF) = \sF(M).$  We note that this property is independent of the
covering.

Sheaves differ from vector bundles mainly due to the fact that the abelian
groups involved, i.e. $\sF(U),$ need not be free.  In the cases we will study,
all our sheaves will be $\cal O$-modules.  Thus, the sections admit
multiplication by locally holomorphic functions.  In this case, and if its
fibers or stalks are finitely generated (the condition is actually slightly
different from this, as we will see), then we can the treat the sheaves as we
would ordinary modules.  As the sheaves are generally not made up of free
abelian groups (and are thus not vector bundles), defining notions similar to
the Euler characteristic will be quite subtle.  To this end, we will briefly
review homological properties of commutative algebra before discussing how to
generalize these concepts to sheaves.

In the following, we will describe point-wise and then global constructions.
So to begin, instead of considering sheaves as $\cal O$-modules, we will
consider modules of the ring
$$O_n \equiv \lim_{\{ 0\}\in U}{\cal O}(U),\qquad U\subset {\bf C}^n.$$
Thus $O_n$ is the ring of convergent power series in $z.$  Some facts about
this ring.  It has a unique maximal ideal equal to the (power series of)
functions vanishing at the origin (they clearly are an ideal since
$f(0)=0\Rightarrow f\cdot g(0) = 0$).  The ring $O_n$ is also Noetherian,
meaning all ideals are finitely generated.

The sheaves we will consider are global versions of $O_n-$modules, which for us
will always be finitely generated (as $O_n$-modules; they may be infinite
dimensional vector spaces).  Any $O_n-$module $M$ defines a module of
relations, $R$ as follows.  If $\{m_1,...,m_k\}$ is a set of generators, then
$$R = \{(\lambda_1,...,\lambda_k) : \lambda_1m_1 + ... + \lambda_km_k = 0\}.$$
$R,$ it can be shown, is also finitely generated.  We then have that the
sequence of $O_n$-modules,
$$0\longrightarrow R\longrightarrow O_n^{(k)}\longrightarrow M\longrightarrow
0,$$
is exact, where $O_n^{(k)} \equiv O_n\oplus O_n\oplus...\oplus O_n$ ($k$
times).   The global analogue of finite-dimensionality is the notion of a
coherent sheaf.  A coherent sheaf is one which has a local presentation
$$\O^{(p)} \longrightarrow \O^{(q)} \longrightarrow \sF \longrightarrow 0.$$
By ``local presentation,'' we mean that for each point $p$ there exists a
neighborhood $U\ni p$ such that the above sequence is an exact sequence of
modules when restricted to $U.$  The $\O^{(q)}$ means that $\sF\vert_U$ is
finitely generated (not just the stalk $\sF\vert_p$) and the $\O^{(p)}$ means
that there are a finite number of relations among these generators.  The gist
of this definition is that it allows us to carry properties of
finite-dimensional modules over to sheaves.

Of course we have the usual properties of modules.  Given two $O_n-$modules $M$
and $N,$ we can construct the $O_n-$modules $M\oplus N,$ $M\otimes_{O_n}N,$ and
$\hom_{O_n}(M,N).$  Note that tensoring and $\hom$ do not necessarily preserve
exactness.
We have instead the following.  Given the exact sequence
$$0\longrightarrow P\longrightarrow Q\longrightarrow R\longrightarrow 0$$
of $O_n-$modules, and an $O_n-$ module $M,$ we have the following exact
sequences:
\eqn\sequences{\matrix{P\otimes_{O_n}M \longrightarrow Q\otimes_{O_n}M
\longrightarrow R\otimes_{O_n}M \longrightarrow 0 \cr
\cr 0\longrightarrow\hom_{O_n}(M,P)
\longrightarrow\hom_{O_n}(M,Q)
\longrightarrow\hom_{O_n}(M,R).} }  The maps are the obvious ones: e.g., if
$\varphi : P\rightarrow Q$ then $\widetilde{\varphi} :
\hom_{O_n}(M,P)\rightarrow \hom_{O_n}(M,Q)$ sends $f$ to $\varphi\circ f.$  The
operations (functors) of $\otimes_{O_n}M$ and $\hom_{O_n}(M,*)$ are said to be
right exact and left exact, respectively.
Na\"\i vely, we would expect the sequences in \sequences\ to extend to a short
exact sequence, and indeed this is the case if the module $M$ is projective
($\Leftrightarrow$ free; we discuss such modules shortly).  The same functors
apply in the category of sheaves, as well.

It is instructive to study how these functors can fail to be exact.  This
situation can arise when the module $M$ -- or sheaf in the global case -- is
not locally free.  Consider the ideal $I\subset O_1$ generated by $z^m.$  Then
we have the exact sequence of $O_1$-modules
\eqn\relseq{0\longrightarrow I\matrix{ {}_{i}\cr\longrightarrow\cr{}}
O_1\matrix{ {}_{\pi}\cr\longrightarrow\cr{}} O_1/I \longrightarrow 0.}
Clearly $O_1/I$ is generated by $\{1,z,...,z^{m-1}\}$ and is an $O_1$ module in
the obvious way. Let us now apply $\otimes_{O_1} O_1/I$ to this sequence to get
$$I\otimes_{O_1} O_1/I\matrix{ {}_{\widetilde{i}}\cr\longrightarrow\cr{}}
O_1\otimes_{O_1} O_1/I\matrix{
{}_{\widetilde{\pi}}\cr\longrightarrow\cr{}}O_1/I \otimes_{O_1}
O_1/I\longrightarrow 0.$$  If this is not exact, then $\widetilde{i}$ has a
kernel.\footnote{$^*$}{The surjectivity of the map $\widetilde{\pi}$ follows
from surfectivity of $\pi.$ For $a\otimes b\in O_1/I \otimes_{O_1} O_1/I,$ we
have $a = \pi(p)$ for some $p$ and thus $a\otimes b = \widetilde{\pi}(p\otimes
b).$  The property is clearly true in general.}  Let us enumerate the
generators.  $\{z^m\otimes z^j, \>j<m\}$ generate $I\otimes_{O_1}O_1/I,$ since
$z^{m+a}\otimes z^b \sim z^m\otimes z^{a+b} = 0$ for $a+b \geq m.$  Note that
we cannot move the $z^m$ across the tensor product since $z^m$ cannot be
written as something in $I$ times something in $O_1$ other than a
scalar.\footnote{$^{**}$}{Recall the definition of tensor product of
$R$-modules:  $A\otimes_R B \equiv (A\times B)/I,$ where $I$ is the ideal
generated by elements $(ra,b)-(a,rb).$} The generators for
$O_1\otimes_{O_1}O_1/I$ are $\{1\otimes z^j,\>j<m\}$ since now any $z$s on the
left side can be moved over by the tensor equivalence.  Under the map
$\widetilde{i},$ $z^m\otimes z^j$ is mapped to $z^m\otimes z^j \sim z^m\cdot
1\otimes z^j \sim 1\otimes z^{m+j} = 0.$  Hence the map $\widetilde{i}$ is
trivial, and $\ker(\widetilde{i}) = I\otimes_{O_1}O_1/I.$  The nontriviality of
the kernel clearly had to do with the torsion of $O_1/I,$ or the existence of
zero divisors.  It is also instructive to check exactness at the middle of this
sequence, and to find the non-surjectivity of the right hand of \sequences\
side when we apply $\hom(O_1/I,*)$ to the original sequence \relseq.  In what
follows, we will define modules which measure the non-exactness of the functors
$\otimes M$ and $\hom(M,*).$

Important to us will be the notions of projective modules and projective
resolutions.  A projective module, $P,$ is one for which the diagram \projdiag\
holds.  That is, given surjective maps $f$ and $g,$ there is a map $h$ such
that $g\circ h = f.$  Another way to put this is that $M\rightarrow
N\rightarrow 0$ is exact implies $\hom(P,M)\rightarrow\hom(P,N)\rightarrow 0$
is also exact.  Thus for projective modules $P,$ $\hom(P,*)$ is a (right and
left) exact functor.  (The sequence could have been extended to a short exact
sequence by adding $\ker(g)$ on the left.)  The same is true of $\otimes P.$
It can be proven that projectivity of a module coincides exactly with its being
free (no relations among the generators, hence isomorphic to a vector space).
In the sheaf language, projectivity is defined similarly, and it coincides with
a sheaf's being {\sl locally} free, i.e. isomorphic to $\O^{(d)}.$

We will use projective modules to ``resolve'' general modules.  In this way,
the complexity of a module will be borne out in the cohomology of the
resolution.  A (left) projective resolution of a module $M$ is an exact
sequence
$$0\longrightarrow P_n\matrix{ {}_{d_n}\cr\longrightarrow \cr {}
}...\longrightarrow P_1\matrix{ {}_{d_1}\cr\longrightarrow \cr {} } P_0\matrix{
{}_{d_0}\cr\longrightarrow \cr {} } M\longrightarrow 0$$
such that each $P_i$ is projective and $d_j\circ d_{j+1} = 0.$  Every
finite-dimensional module has a projective resolution.  One builds it
iteratively, beginning with the sequence of relations \relseq\ and continues to
use \relseq\ on the kernel of the left-most term in the resolution.

When discussing sheaves, we call a resolution by locally free sheaves a syzygy.
 Coherent sheaves have syzygies for the same reasons as above.  Given a syzygy
of some sheaf $E,$ we can analyze the complexity of the sheaf, loosely
speaking, by measuring the extent to which $\hom(E,*)$ fails to be exact, for
example.  To do this we define $\ext$ as follows, first for modules.  Given
$O$-modules $M, N,$ we can first construct a resolution $\{ P_i \}$ of $M$ then
create the complex $C^\cdot$:
\eqn\extseq{0\longrightarrow
\hom(P_0,N)\longrightarrow
\hom(P_1,N)\longrightarrow
\hom(P_2,N)\longrightarrow ...}
(not exact).  We then define
$$\ext^i(M,N) = H^i(C^\cdot)$$
To define the vector spaces $\ext^i(\sF,\sG)$ (which can be thought of as
trivial sheaves) we first resolve the sheaf $\sF$ with a syzygy $\{ {\cal P}_i
\}.$  We note that the stalk of the sheaf, $\sF_p,$ at a point $p$ is nothing
other than an $O$ module; likewise for $\sG.$  The sheaf
$\underline{\ext}^i(\sF,\sG)$ has the natural property that
$\underline{\ext}^i(\sF,\sG)_p = \ext^i(\sF_p,\sG_p)$ and is built by applying
$\ext$ to the local syzygy.  The global version, $\ext^i(\sF,\sG),$ must be
defined more carefully.  We have a complex of sheaves $\{ \hom({\cal P}_i,\sG)
\},$ analogous to \extseq, which gives rise to a double complex.  The
horizontal differential is just the natural one from the complex, while the
vertical direction is defined by the \u Cech cohomology of the sheaves,
reviewed in this section.  $\ext^i(\sF,\sG)$ is the $i^{\rm th}$ term of the
cohomology of the total complex of this double complex.\footnote{$^*$}{Given a
double complex $C^{ij},\> i,j >0,$ with horizontal differential, $d$ and
vertical differential $\delta$ obeying $d\delta + \delta d = 0$ (if $d$ and
$\delta$ commute, we can redefine the signs of odd $d$'s so the two
anti-commute), we define the total complex $C^n = \oplus C^{n-i,i}$ with
differential $D = d + \delta.$}  We note without proof that if
$\underline{\ext}^q(\sF,\sG) = 0$ for $q<k$ then $\ext^k(\sF,\sG)$ is equal to
the global sections of $\underline{\ext}^k(\sF,\sG).$

\subsec{Exceptional Collections and Helices}
A coherent sheaf $E$ over a variety $M$ is called {\sl exceptional} if
$$\eqalign{&\ext^i(E,E) = 0, \qquad\; i\geq1,\cr
&\ext^0(E,E) \cong{\bf C}, \qquad i=0.}$$
These conditions imply that $E$ is locally free, i.e. we can think of $E$ as
the sheaf of sections of some vector bundle.  We sometimes refer to $E$ itself
as a vector bundle.

An ordered collection of exceptional bundles $\epsilon = (E_1, ..., E_k)$ is
called an exceptional collection if
for all $1\leq m<n\leq k$ we have
$$\eqalign{&\ext^i(E_m,E_n) = 0, \qquad i\geq1,\cr
&\ext^i(E_n,E_m) = 0, \qquad i=0.}$$

The most important property that exceptional collections enjoy is that they can
be transformed to get new exceptional collections.  We will define right and
left transformations or mutations.  Together, they will represent an action of
the braid group on the set of possible such collections.

Mutations of collections by the action of the braid group are performed by
making replacements of neighboring pairs.  Take a neighboring pair
$(E_i,E_{i+1})$ of sheaves in an exceptional collection, $\epsilon =
(E_1,...,E_n).$  Suppose the following condition is true:
$\hom(E_i,E_{i+1})\neq 0$ and
$$\hom(E_i,E_{i+1})\otimes E_i\matrix{ {}_{\rm ev}\cr \longrightarrow\cr {}}
E_{i+1}\longrightarrow 0$$
is exact (i.e., ${\rm ev}$ is surjective), where the map ${\rm ev}$ is the
canonical one.  Then we can define a new sheaf $L_{E_i}E_{i+1}$ to be the
kernel of this map:
$$0\longrightarrow L_{E_i}E_{i+1} \longrightarrow \hom(E_i,E_{i+1})\otimes
E_i\longrightarrow E_{i+1}\longrightarrow 0.$$  For brevity, we usually write
this new sheaf as $LE_{i+1}.$  Thus we write $L^2 E_{i+1}$ for
$L_{E_{i-1}}(L_{E_i}E_{i+1}),$ etc.  If we then replace the pair
$(E_i,E_{i+1})$ by $(LE_{i+1},E_i)$ in the exceptional collection, we have the
following.

{\bf Theorem:} The new collection $\epsilon^\prime =
(E_1,...,E_{i-1},LE_{i+1},E_i,E_{i+2},...,E_n)$ is exceptional.

Sometimes the canonical map $\hom(E_i,E_{i+1})\otimes E_i\rightarrow
E_{i+1}\rightarrow 0$ is not surjective.  If, however, it is injective, we can
define $LE_{i+1}$ to be the cokernel of this map instead of the kernel, and the
theorem is still true.  Finally, if $\hom(E_i,E_{i+1}) = 0$ but
$\ext^1(E_i,E_{i+1})\neq 0,$ we can define the left mutation $L_{E_i}E_{i+1}$
to be the universal extension, defined by its property of making the sequence
\eqn\leftmut{0\longrightarrow E_{i+1}\longrightarrow L_{E_i}E_{i+1}
\longrightarrow \ext^1(E_i,E_{i+1})\longrightarrow 0}
exact.

To demonstrate when ${\rm ev}: \hom(A,B)\otimes A\rightarrow B$ is not
surjective, consider the simple example of modules (not sheaves) on the complex
plane, ${\bf C}.$  Define $A = \O/I_2,$ $B = \O/I_4,$ where $I_m$ is the ideal
generated by $z^m.$  Then $\{1,z\}$ generate $A$ and $\{1,z,z^2,z^3\}$ generate
$B.$  Now $f\in\hom_{\O}(A,B)$ must satisfy $f(z^na) = z^nf(a),$ and since $z^2
=0$ in $A,$ we have $z^2f(a) = 0$ in $B.$  Thus, $f(1) = c_1z^2 + c_2z^3,\,
c_i\in {\bf C}.$  Clearly, the image of ${\rm ev}$ is ${\bf C}z^2\oplus{\bf
C}z^3,$ which is not all of $B.$

We can perform right mutations as well, under different conditions.  Since
$\hom(E_i,E_{i+1})$ as a vector space has the identity morphism in
$$\hom((\hom(E_i,E_{i+1}),\hom(E_i,E_{i+1})) \cong
\hom(E_i,E_{i+1})^*\otimes\hom(E_i,E_{i+1}),$$ we get a canonical map from
$E_i$ to $\hom(E_i,E_{i+1})^*\otimes E_{i+1}.$  To understand this, it may help
to consider the finite-dimensional vector space example.  Choose bases
$a_i,b_j$ for $E_i,E_{i+1}$ and dual bases $\widetilde{a}^i,\widetilde{b}^j$
for $E_i^*,E_{i+1}^*.$  Then the map sends $a
\mapsto\sum_{j}(a\otimes\widetilde{b}^j)\otimes b_j.$  If this map is
injective, then the right mutation is then defined by the cokernel as shown:
\eqn\rightmut{0\longrightarrow E_i\longrightarrow \hom(E_i,E_{i+1})^*\otimes
E_{i+1}\longrightarrow R_{E_{i+1}}E_i\longrightarrow 0.}  Similarly to the left
mutation, we can define the right mutation if this map is surjective, or if
$\ext^1(E_i,E_{i+1})\neq 0.$

We note here -- though it should not be seen as obvious -- that if the
exceptional pair $(A,B)$ admits a left mutation, then the pair $L(A,B) =
(LB,A)$ is exceptional (by the theorem above) and its right mutation is such
that $R(LB,A) = (A,B).$  Thus, if the left operation is seen as a braiding (we
have yet to show this), then the right mutation is an unbraiding.  The right
operation is dual in the following sense.  If $\epsilon = (E_1,...,E_n)$ is
exceptional then so is $\epsilon^* \equiv (E_n^*,...,E_1^*),$ and if $\epsilon$
admits a left transformation abbreviated $L\epsilon,$ then $\epsilon^*$ admits
a right transformation.  We then have $(L\epsilon)^* = R\epsilon^*.$ If
$R\epsilon$ exists then so does $L\epsilon^*,$ and in that case $(R\epsilon)^*
= L\epsilon^*.$

These mutations amount to an action of the braid group of $n$-objects on
helices, as they obey the Yang-Baxter relations.  Specifically, if $L_i$
represents mutating $(E_i,E_{i+1})$ to $(L_{E_{i}}E_{i+1},E_i)$ -- i.e. a shift
at the $i^{\rm th}$ entry of the foundation -- then
\eqn\YB{L_iL_{i+1}L_i = L_{i+1}L_iL_{i+1}}
is satisfied.  The same is true of the right shifts.

A {\sl helix} is an infinte collection of coherent sheaves $\{E_i\}_{i\in{\bf
Z}}$ such that
$$\eqalign{&1)\> \hbox{for any } i\in {\bf Z}, (E_{i+1},...,E_{i+n})\> \hbox{is
an exceptional collection.}\cr & 2)\> R^{n-1}E_i = E_{i+n.}}$$
Thus when you move $n$ steps to the right, you come back to where you were, up
to a translation.  This explains the terminology.  It is clear that a helix is
uniquely determined by any of its foundations, and conversely any exceptional
collection determines a helix.

We are finally in a position to define the bilinear form which corresponds to
the matrix $S$ of soliton numbers.  Let us assume we have a helix $\epsilon$
with foundation $(E_1,...,E_n).$  In the case of bundles, this form is the
relative Euler characteristic, and is defined generally to be
\eqn\bilin{\chi(E_i,E_j) \equiv \sum_{k = 0}^{n}(-1)^k {\rm
dim}\ext^k(E_i,E_j).}
We took care to define this form even for sheaves which do not correspond to
bundles.  For vector bundles, though, computations are simplified by the
following derivation.
Since these sheaves $E_i$ are locally free (they are sections of a vector
bundle), the syzygy or projective resolution is trivial:  it has the form
$0\rightarrow P_0\rightarrow E_i\rightarrow 0,$ with $P_0 = E_i.$ Since the
complex $\{ P\}$ just contains a single element , the double complex used to
compute $\ext$ reduces to a single complex, and $\ext(E_i,E_j)$ becomes the
ordinary sheaf cohomology of $\hom(E_i,E_j).$  Thus
$${\rm dim}\,\ext^k(E_i,E_j) = {\rm dim}\, H^k(\hom(E_i,E_j))$$ and so
$$\eqalign{\chi(E_i,E_j) &= \sum_{k}(-1)^k{\rm dim}\,H^k(\hom(E_i,E_j)) \cr
&= \sum_{k}(-1)^k{\rm dim}\,H^k(E_i^*\otimes E_j)\cr
&= \chi(E_i^*\otimes E_j) \cr
&= \int_{M}({\rm ch}(E_j)/{\rm ch}(E_i)){\rm td}(TM,\cr}$$
where in the second line we considered the sheaves as vector bundles, using the
equivalence of sheaf and vector-valued cohomology (the de Rham theorem).  The
last line expresses the Euler characteristic in terms of characteristic classes
of bundles in de Rham cohomology, and is equivalent to the Riemann-Roch
theorem.

\subsec{Examples:  Projective Spaces, Grassmannians, Orbifolds, and Blow-Ups}

The simplest spaces for us to consider are the projective spaces.  We recall
the connection between divisors and line bundles.  A hyperplane is described by
the zero locus of a linear polynomial in homogeneous coordinates (which is
itself a section of a line bundle).  Thus, up to isomorphism, all hyperplanes
have the form $X_0 = 0.$  To find the corresponding line bundle, look at the
set $U_1 = \{X_1\neq 0\}.$  Then the hyperplane $H$ is described by $z_0 \equiv
X_0/X_1 = 0.$  In $U_2 = \{X_2\neq 0\},$ it is given by $w_0 \equiv X_0/X_2 =
0,$ and on the intersection the two functions are related by a nonzero
function: $z_0 = (X_2/X_1)w_0,$ which defines a transition function between
open sets.  The set of all such transition functions determines a line bundle.
More generally, a divisor defined by a function $f_\alpha$ in $U_\alpha$ and
$f_\beta$ in $U_\beta$ defines holomorphic transition functions
$s_{\alpha\beta}$ on $U_\alpha\cap U_\beta$ by $$f_\alpha =
s_{\alpha\beta}f_\beta.$$  Different defining functions for equivalent divisors
yield isomorphic line bundles.  We find that the isomorphism classes of line
bundles are given by the degrees of the defining polynomials (with negative
degrees associated to divisors along poles), and so all line bundles can be
written as powers of the hyperplane bundle described above.  We denote the
$d^{\rm th}$ power of the hyperplane bundle by ${\cal O}(d).$

{\bf Theorem:}  The collection $\{{\cal O}(m)\> :\> m\in{\bf Z}\}$ of sheaves
on ${\bf P}^n$ is exceptional, and $({\cal O},{\cal O}(1),...,{\cal O}(n))$ is
a foundation of the helix.  We define $E_i \equiv {\cal O}(i).$

To show that this space is a helix, we have to study the mutations.  Here we
will just consider the first right mutation.  Consider the following exact
sequence of sheaves:
$$0\longrightarrow \O \longrightarrow V\otimes \O(1) \longrightarrow T
\longrightarrow 0.$$
Here $V$ is the $n+1$ dimensional vector space of which ${\bf P}^{n}$ is the
projectivization (the notation $V$ above thus signifies the trivial rank $n+1$
vector bundle) and $T$ is the holomorphic tangent space.  This sequence is
called the Euler sequence.  Locally, for some choice of basis for $V,$ sections
of $V\otimes \O(1)$ look like $(f_0,...,f_n)$ and are mapped to the vector
field $\sum_i f_i \del_i$ (easily checked to be well-defined when the $f_i$ are
in $\O(1)),$ where we have the relation $\sum_i X_i\del_i = 0.$  The
one-dimensional kernel is the image of the left map.  We need to show that the
maps are the canonical ones of \rightmut\ and that $V\cong\hom(\O,\O(1))^*.$
We show only the latter.  In fact, this is readily checked.  The notation
$\hom$ stands for the global sections, here, and since $V\cong V^*$ as vector
spaces, we need to show ${\rm dim} H^0\hom(\O,\O(1)) = n+1.$  Since $\hom$ is
taken over $\O,$ we can identify $\hom(\O,\O(1))\cong\O(1).$  The global
sections of $\O(1)$ are linear functionals on $V,$ and there are $n+1$ of them
-- the $n+1$ coordinate functions, for example.Thus, the above sequence gives
the right mutation and $R\O = T$

Further right mutations (by ``exterior products'' of this sequence) show that
as we move $\O$ right by mutations we find $R^k\O = \Lambda^kT,$ and indeed
$R^n\O = \Lambda^nT = \O(n+1),$ as it must \rGor.

Now that we have an exceptional collection of vector bundles, we may compute
the bilinear form
$$\chi(E_i,E_j) = \chi(E_i^*\otimes E_j).$$
The Todd class and the Chern characters of vector bundles are computed in terms
of the Chern roots -- two-forms computed from the splitting principle, or
diagonalizing the curvature.  For  the tangent bundle ${\bf TP}^n,$ each Chern
root $x_i$ is equal to $x,$ the K\"ahler two-form.  The Todd class is given by
$${\rm td}(E) \equiv \prod_{i=1}^{r}{x_i\over 1-{\rm e}^{-x_i}} = {x^{n+1}\over
(1-{\rm e}^{-x})^{n+1}}\qquad \hbox{for $T{\bf P}^n.$}$$  The Chern character
is defined to be ${\rm ch}(E) = \sum {\rm e}^{x_i}.$  We have $${\rm
ch}(\O_{{\bf P}^n}(m)) = {\rm e}^{mx}.$$
Therefore,
$$\eqalign{\chi(E_i,E_j) = \chi(E_i^*\otimes E_j)&= \chi(\O(i)^*\otimes
\O(j))\cr &= \chi(\O(j-i))\cr &= \int_{{\bf P}^n}{\rm ch}{(\O(j-i))}{\rm
td}(T{\bf P}^{n}) \cr &= \int_{{\bf P}^n} {\rm e}^{(j-i)x}{x^{n+1}\over (1-{\rm
e}^{-x})^{n+1}}.}$$
The integrand is understood as a polynomial in the two-form $x.$  For ${\bf
P}^n,$ we have $\int_{{\bf P}^n} x^n = 1,$ so we simply need to extract the
coefficient of $x^n.$  To do this, we multiply by $x^{-n-1}$ and take the
residue:
$$\eqalign{\chi(E_i,E_j) &= \oint {{\rm e}^{(j-i)x}\over  (1-{\rm
e}^{-x})^{n+1}} dx \cr
&= \oint (1-y)^{-(j-i)-1}y^{-n-1} dy,\qquad y = 1-{\rm e}^{-x} \cr
&= {1\over n!}\left({{\rm d}\over {\rm
d}y}\right)^n(1-y)^{-(j-i)-1}\vert_{y=0}\cr
&= {1\over n!}((j-i)+1)((j-i)+2)...((j-i)+n)\cr
&= \pmatrix{n+(j-i)\cr (j-i)}.}$$
Indeed, this is related to the result we obtained for the (weighted) soliton
numbers of the topological sigma model on ${\bf P}^n.$  Specifically, we have
found the inverse of the matrix \pnnums.  As we discussed in section 2.4, a
matrix and its inverse are equivalent under braiding.  Therefore, the
conjectured math-physics link has been demonstrated.  What is more, the
fundamental solitons of the physical theory were shown to be given by the
coordinate functions of the ${\bf C}^{n+1}$ which fibers over ${\bf P}^n.$
These are none other than the global sections of the bundle $\O (1).$  Note
that
$${\rm dim}\,H^0(\O(1)) = {\rm dim}\,Ext^0(\O(k),\O(k+1)) =
\chi(\O(k),\O(k+1)).$$
Here, then, the physical and mathematical calculations are counting the same
things!  Unfortunately, the correspondence does not seem so direct for other
examples.

Exceptional collections for Grassmannians and other flag manifolds were
considered by M. M. Kapranov in \rKapranov\ and in the sixth article in
\rHelices.  These authors were able to construct exceptional collections by
relating vector bundles over homogeneous spaces to representations of the coset
group in the standard way.  Namely, on $G/H$ we can define a vector bundle
given any representation of $H$ by the associated bundle of the principal $H$
bundle $G \rightarrow G/H.$\footnote{$^*$}{Given a principal $H$ bundle, $G,$
with transition functions $h_{\alpha\beta}$ and a representation $\rho:
H\rightarrow {\rm Aut}(V)$ of $H,$ we construct the associated rank $n = {\rm
dim}V$ vector bundle by considering the transition functions $s_{\alpha\beta} =
\rho(h_{\alpha\beta}).$  Equivalently, we take the space $G\times_\rho V$
defined to be $G\times V$ modulo the relation $(gh,v) \sim (g,hv).$}  The
Grassmannian $G(k,N)$ of $k$-planes in ${\bf C}^N$ is equated with
$U(N)/(U(k)\times U(N-k))$ (by considering that $U(N)$ acts transitively on the
set of planes, with $U(k)\times U(N-k)$ fixing a given plane), and is thus a
homogeneous space.  We can take representations of $U(k)$ alone to define
vector bundles.  These representations are described by Young diagrams, where
now we allow negative indices, as the totally antisymmetric representation acts
by the determinant (which is trivial for $SU(k)$ but not for $U(k)$), which can
be raised to any (positive or negative) power.

Kapranov has shown that an exceptional collection is defined by the young
tableaux with the property that all entries are non-negative and no row has
more than $(N-k)$ elements.  We note that there are $\pmatrix{n\cr k}$ such
diagrams, equal in number to the Euler character and the dimension of the
Grothendieck group (the Hodge diamond is diagonal).

In order to compute the bilinear form, we note that the tensor products of
these line bundles are nothing other than the bundles associated to the tensor
products of the representations.  Further, the dual bundle is described by the
dual representation, defined by reversing the sign and order of the Young
tableau indices (thus $(\alpha_1,...,\alpha_k)^* = (-\alpha_k,...,-\alpha_1)$).
 Therefore, in order to compute the Euler character $\chi(E,F)$ we just
decompose the tensor product $E^*\otimes F$ and take the Euler character of
each component separately.  Kapranov \rKappriv\ has shown that this quantity is
nonzero only when all of the Young indices are non-zero, and equal to the
dimension of the representation (as a representation of $U(N)$) in this case.
Therefore, we have reduced the problem of computing this bilinear form to a
question of representation theory.

Note, too, that there is a partial ordering on the representations in terms of
inclusion of Young diagrams.  As $E^*\otimes F$ contains positive parts only
when $E$ appears as a subdiagram of $F,$ the upper-triangularity of the
bilinear form follows immediately.  These results are readily extended to the
flag manifolds $U(N)/(U(n_1)\times...\times U(n_r)),$ $\sum n_j = N.$  See Ref.
\rKapranov.

Let us compute the first nontrivial example: the Grassmannian $G(2,4) \cong
U(4)/(U(2)\times U(2)).$  The basis $\{e_1,...,e_6\}$ for the bundles is given
by the diagrams
$$(0,0),(1,0),(2,0),(1,1),(2,1),(2,2),$$
which we label by $e_i,\> i = 1...6.$
where the $i^{\rm th}$ entry denotes the length of the $i^{\rm th}$ row.  Note
that the totally anti-symmetric $(1,1)$ diagram, for example, is trivial as a
representation of $SU(2)$ but is in fact the one-dimensional representation in
which a matrix in $U(2)$ acts by its determinant.  We will do a sample
computation of a matrix element in the bilinear form.  Consider $\chi_{2,5} =
\chi(e_2,e_5).$  We need to decompose $$e_2^*\otimes e_5 = (0,-1)\otimes (2,1)
= ((-1,-1)\otimes (1,0))\otimes ((1,1)\otimes (1,0)) = (1,0)\otimes (1,0).$$
where in the last steps we have factored out the $({\rm det})^{\pm1}$
representations (which cancel when tensored).  Using the usual rules of
decomposition, we find $(1,0)\otimes (1,0) \cong (2,0)\oplus (1,1),$ where the
second summand not trivial in $U(2).$  We find
$\chi_{2,5} = {\rm dim}_{U(4)}(1,1) + {\rm dim}_{U(4)}(2,0),$
where the subscript indicates that we take the dimension of the representation
as a representation of $U(4).$  Hence $\chi_{2,5} = 10 + 6 = 16.$
Proceeding straightforwardly, we find
$$\chi = \pmatrix{1&4&10&6&20&20\cr
0&1&4&4&16&20\cr 0&0&1&0&4&10\cr 0&0&0&1&4&6\cr
0&0&0&0&1&4 \cr 0&0&0&0&0&1}.$$
As it must, the matrix $\chi$ satisfies ${\rm det}(z - \chi\chi^{-t}) =
(z-1)^6.$
Other Grassmannian spaces -- including projective spaces -- can be computed
similarly.  We note that $G(N-1,N),$ equivalent to $G(1,N) = {\bf P}^{N-1},$
gives a basis for which the bilinear form is the same as in the physics case
(no matrix inversion is necessary).

The case of orbifolds, discussed from the physical viewpoint in section 2.5,
has not been studied from the mathematical viewpoint in the context of helices.
 However, Bondal and Kapranov have discussed the derived category of complexes
of ${\bf Z}_3$-equivariant sheaves over ${\bf P}^1,$ and have found results
similar to those derived in section 2.5 \rBK.  The sheaves are no longer
locally free, but include those associated to the fixed points of the ${\bf
Z}_3$ action.  Similar results hold for any finite group, leading us to
conjecture that the bilinear forms will be associated to Dynkin diagrams, as
was previously discussed.

It is interesting to consider blow-ups of ${\bf P}^2$ at $n$ points.  We will
encounter an example of such a space (when $n = 1$) in some detail in section
4.4.  For now, we just note that these spaces, which we shall denote
$\widetilde{{\bf P}^2_n},$ have isomorphisms
$$\widetilde{{\bf P}^2_n} - \cup_{i=1}^{n}E_i \cong {\bf P}^2 -
\cup_{i=1}^{n}\{p_i\}.$$
That is, each point $p_i, \; i= 1...n$ is replaced by an ``exceptional
divisor'' $E_i,$ isomorphic to ${\bf P}^1.$  The isomorphism arises from a map
$\pi:\widetilde{{\bf P}}^2_n\rightarrow{\bf P}^2,$ with $\pi^{-1}(p_i) = E_i.$

An exceptional collection can be defined for these spaces, though whether they
generate helices or not is not known (for most examples; for $\widetilde{{\bf
P}}^2_1,$ see section 4.4).  The collection is given by ${\cal O}(i),\;
i=0...2$ -- sections of the $i^{\rm th}$ power of the hyperplane bundle, pulled
back under $\pi$ -- and the sheaves ${\cal O}_j$ $j=1...n,$ which have support
only on $E_j.$  That is ${\cal O}_j(U) = {\cal O}(U)/I(U),$ where $I$ is the
ideal of holomorphic functions vanishing on $E_j.$  Clearly, ${\cal O}_j(U) =
0$ for $U\cap E_j = \emptyset.$  One finds for the bilinear form (say, for
$n=3$) \rPol:
$$\chi = \pmatrix{1&3&6&1&1&1\cr
0&1&3&1&1&1\cr 0&0&1&1&1&1\cr 0&0&0&1&0&0\cr
0&0&0&0&1&0\cr 0&0&0&0&0&1}.$$

We note that $H = \chi\chi^{-t}$ is unipotent, yielding the correct charges
(Hodge numbers) for the topological sigma model, including the integer parts
(see ${\bf P}^1$ example in section 2.5).  Namely, the eigenvalues of $H,$ all
equal to unity, wrap around the origin an integral number of times when $\chi$
is given a $t$ dependence and is taken from the identity to $\chi$ smoothly.
The integer value is precisely the form degree.  The Hodge numbers of the space
are determined as follows:  each exceptional divisor, isomorphic to ${\bf
P}^1,$ contributes one $(1,1)$ form, in addition to the ${\bf P}^2$ cohomology
which pulls back from $\pi;$  thus $h^{ij} = {\rm diag}(1,4,1),$ which is what
is found.

We note that if $n>8,$ then these spaces have negative first Chern classes, and
therefore do not have a simple geometric interpretation as asymptotically free
quantum field theories.  Nevertheless, the mathematical constraints are
satisfied.  The difference may arise from the difference between an exceptional
collection and a helix:  admissibility of mutations and the `periodicity'
requirement of the shifts $R^n$ and $L^n.$

\newsec{Links}
\subsec{Parallel Structures, Categorical Equivalence}

One way of formalizing the parallel structures shared by topological field
theories and exceptional collections is by describing a categorical equivalence
between the two.  In fact, one approach to the theory of helices is through
their categorical definition.  Many of the structures we have discussed are
structures which can be defined given an abelian category and its derived
category.

For those of us unfamiliar with categorical constructions, we recall only the
very basics.  One constructs a category of objects (e.g., sets, topological
spaces, sheaves, groups, vector spaces, complexes) and composable morphisms
(e.g., functions, continuous functions, maps of sheaves, homomorphisms, linear
maps, morphisms of complexes).  Categories may have additional structures such
as addition of objects (e.g., direct sums of vector spaces or complexes).
Functors are maps between categories which map objects to objects and morphisms
to morphisms, respecting composition.  For example, the fundamental group
$\pi_1$ is a functor from topological spaces to groups.  Continuous functions
are mapped to group homomorphisms.

Equivalence of two categories, $\cA$ and $\cB,$ is provided by constructing a
bijective functor -- an invertible recipe for getting $\cB$ objects and maps
from such $\cA$ structures.  First, let us consider the example of the category
of coherent sheaves.  As every coherent sheaf $\sF$ has a syzygy
$\sP_n\rightarrow...\sP_0$ with homology $H_0(\sP^\cdot) = \sF,$ we may focus
our attention on the category of complexes of sheaves, defined up to
(co)homology, instead of just the category of sheaves.  This is called the
derived category of coherent sheaves.

We now pass from sheaves to algebras using the following construction \rBondal.
 Consider a sheaf, $\sF,$ an object of the category $\cA$ of coherent sheaves.
Then we have an algebra $A = \hom(\sF,\sF),$ and we can use $\sF$ to construct
the map (functor)
$$F_\sF(\sG) = \ext^\cdot(\sF,\sG)$$
from the category of sheaves to (bounded) complexes of representations of $A,$
or $D^\flat({\rm mod}-A).$  Here the differential map on the complex of
$\ext$'s is the zero map.  Of course the interesting question is when is this
functor a category equivalence.  A. I. Bondal has shown that when $\sF$
includes sufficiently many summands -- usually, $\sF$ will be a direct sum of
generators for $\cA$ -- then this is so \rBondal.

This statement is profound.  It allows us to shift our focus to
finite-dimensional algebras.  What is more, we can analyze the algebraic
properties associated to exceptional collections and helices.  What are the
special algebras and representations which result from this construction?
The algebras turn out to be those associated to {\sl quivers,} which we will
briefly review here.

[Before discussing quivers, it should be noted that helices can be defined not
only for collections of sheaves, but for an arbitrary triangulated category (a
triangulated category is an abelian category ${\cal C},$ with an automorphism
functor $T: {\cal C}\rightarrow{\cal C}$ which satisfies certain axioms; the
paradigm is the category of complexes with automorphism a shift in \u Cech
degree).  We will not discuss these matters in detail, except to say the the
constructions are rather general.  We feel the geometric link to physics is of
primary importance, and have therefore concentrated our attention on this
application of helix theory.  For more on the categorical approach to helices,
see the first and seventh articles in Ref. \rHelices.]

A quiver is a set of labeled points with some number of labeled directed arrows
between them.  An ordered quiver is one in which the points are ordered and
arrows connect points labeled with a lesser number to those with a greater
label.  For example, the diagram \quivdiag\ defines a quiver.
Path composition defines an algebra, $A,$ associated to a quiver, with vertices
$p_i$ corresponding to projections in the algebra: $p_i\cdot p_i = p_i,$ and
products equaling zero if paths don't line up tip to tail.  We may impose
relations in the algebra.  In the example above, we may put
\eqn\quivrels{f_i\cdot g_j = f_j\cdot g_i.}
In this case, we say we have a {\sl quiver with relations.}

The projections $p_i$ decompose representations $V$ of $A$ (or left
$A$-modules) into (non-invariant) subspaces $V_i,$ via
$$V = \oplus_{i}p_iV = \oplus_{i} V_i,$$
and arrows determine morphisms $V_i\rightarrow V_j.$  Now suppose $W$ is a {\sl
right} $A$-module.  Then we can also decompose $W$ into subspaces $G_iW \equiv
Wp_i$ via
$$W = \oplus_{i}Wp_i = \oplus_{i} G_iW.$$  We can then consider $A$ as a right
module over itself and define submodules $P_k = p_kA$ (closed under $A$).  Now
the algebra of a quiver  with $n+1$ vetrtices looks like $A = \oplus_0^nP_i,$
which is naturally identified with
$$A = \hom_A(A,A) = \hom_A\left(\oplus_{i=0}^{n}P_i,\oplus_{j=0}^{n}P_j\right)
= \oplus_{i,j}\hom(P_i,P_j).$$
Therefore the algebra of a quiver is the algebra of morphisms between modules.
Also note that $\hom(P_i,P_j) = 0$ for $i>j,$ if the quiver is ordered.

Suppose we have a strong exceptional collection, by which we mean an
exceptional collection satisfying the additional requirement that
$\ext^k(E_i,E_j) = 0$ for all $i$ and $j$ when $k\neq 0.$  (As stated, we will
work here with sheaves over $X.$)  Then if the derived category is generated by
the collection, we can write $E = \oplus E_i$ and define $A = \hom(E,E).$  Then
$A$ is the algebra of a quiver with relations, and Bondal has proven that this
mapping from sheaves to right $A$-modules is an equivalence of categories:
$$D^\flat({\rm Sheaf}(X) \cong D^\flat({\rm mod}-A).$$

An example of a strong exceptional collection is $(\O,\O(1),...,\O(n))$ over
${\bf P}^n.$  Consider $n=2.$  Then $$A =
\hom(\O\oplus\O(1)\oplus\O(2),\O\oplus\O(1)\oplus\O(2)),$$
which is the algebra of the quiver \quivdiag\ with the relations \quivrels.
The vertices $p_i$ correspond to the one-dimensional spaces
$\hom(\O(i),\O(i)),$ and the arrows correspond to the three independent
generators of $\hom(\O(i),\O(i+1))$ (see calculation for ${\bf P}^n$ in section
3.4), which compose according to the quiver relations.  Further, mutations act
similarly on exceptional collections and modules\footnote{$^*$}{We define the
(right) mutation of representation spaces $R(V_i,V_{i+1}) = (V_i^\prime,
V_{i+1}^\prime)$ by $V_i^\prime = V_{i+1}$ and $V_{i+1}^\prime= {}^\bot
V_{i+1}\cap V_i\oplus V_{i+1}.$}

We remark here that not all exceptional collections have all mutations
admissible, nor is it known whether a strong exceptional collection can always
be found.  Further, the helix generated by an exceptional collection might not
yield a quasi-unipotent bilinear form (if $R^nE$ were not represented by
tensoring by a line bundle, for example).

Another notion which these remarks cannot address is the question of
constructivity.  Can any exceptional collection be generated by mutations of a
given one?  For ${\bf P}^1$ and ${\bf P}^2,$ for example, it has been proven by
Drezet and Rudakov that this is so.  For ${\bf P}^2,$ the conditions of being
exceptional yield the Markov equation, $x^2+y^2+z^2-3xyz = 0,$ for the ranks of
the bundles involved \rRud.  For higher dimensional spaces, little more is
known.  Note that the Markov equation is precisely the equation of
classification for topological sigma models with three vacua (see section 6.2
of Ref. \rCVNEW).  The constructivity of the helix then tells us that this is
the only such model with three vacua (up to continuous deformations, as
always).  It would be interesting to see these results translated into the
simple -- or more intuitive -- structure of quivers.

\subsec{Localization}

The difficulty of the classification program is that finding the soliton
numbers of a physical theory is a daunting task, especially in the sigma model
case.  One must first find the quantum ring of the topological sigma model,
which itself demands a detailed knowledge of rational curves on the manifold.
Then, one must construct and solve the $tt^*$ equations.  Very few solutions to
these equations are known.  At best, some asymptotics have been calculated in a
limited number of cases.  This is what is needed in order to extract the
soliton numbers.  These numbers may be calculable numerically, but that
program, too, is formidable.  A quick way of deriving soliton numbers would be
a godsend.  This is why a link would be so interesting.

In this section, we shall outline a possible approach to this problem.  While
difficulties remain, we hope that these obstacles can be overcome.  The idea is
to localize the vacua by adding a potential to the sigma model.  Normally, this
would destroy the N=2 invariance, as the potential would have to be a
holomorphic function of the superfield coordinates (and the only holomorphic
functions on a compact manifold are the constants, which don't affect the
Lagrangian due to the integration over superspace).
However, this difficulty can be circumvented if the manifold admits a
holomorphic Killing vector for its K\"ahler metric.  We recall that any
manifold admits a supersymmetric sigma model -- a natural extension to
superfields of the ordinary nonlinear sigma model.  If the manifold is K\"ahler
with metric $g_{i\oo{j}},$ then the separation into holomorphic and
anti-holomorphic tangent spaces (preserved by the connection) leads to a second
independent supersymmetry.  A potential is added to this model by inserting a
more general form for the Lagrangian, accompanied by new transformation laws.
Conditions on the potential terms arise from requiring this Lagrangian to be
invariant under supersymmetry.

We report here the results of this procedure, calculated first in \rPot.  Let
us call the Lagrangian in \action\ $S_0(\Phi)$ (here $\Phi$ denotes all of the
fields).  Then the sigma model with potential has the form
\eqn\potact{S_0(\Phi) + m^2g_{\mu\nu}V^{\mu}V^{\nu} + m\oo{\psi}^\mu D_\mu
V_\nu \psi^\nu.}
Here $V$ is the holomorphic Killing vector, which by definition obeys
$$\eqalign{D_\mu V_\nu + D_\nu V_\mu &= 0\cr
\partial_i V_{\oo{j}} + \partial_{\oo{j}}V_i &= 0.}$$
Since $V$ is holomorphic, we have $V^{\oo{i}} = (V^i)^*,$ and the second
condition above tells us that we may write
$$V_a = i\partial_a U,$$
where $U$ is a real (not analytic) function  (the metric relates our
holomorphic Killing vector to a closed one-form, which is therefore the
exterior derivative of a function, as the diagonality of the Hodge numbers
implies $H^1 = 0$).  We may proceed with the topological twisting in the usual
fashion, redefining the bundles of which the fermions are sections.  To obtain
the topological theory, we need the action of $Q_+.$  We find
$$\eqalign{[Q_+,\phi^i] &= i\chi^i\cr \{Q_+,\chi^i \} &= -iV^i \cr
\{Q_+,\rho^i_{\oo{z}} \} &= \partial_{\oo{z}}\chi^i -
i\Gamma^i{}_{jk}\chi^j\rho_{\oo{z}}^k.\cr}$$
Note that if we now try to do the usual game of relating local observables to
differential forms by $\chi^i \leftrightarrow dz^i$ etc., we find that
$Q = d + mi_V \equiv d_m,$
where $i_V$ is contraction by $V$ and $m$ is a parameter which scales with $V$
(we have defined $Q \equiv Q_+$).
It is clear then that
$$Q^2 = m(di_V + i_Vd) = m{\cal L}_V,$$
where ${\cal L}_V$ is the Lie derivative in the $V$ direction (the second
equality above is true for differential forms).  Thus, the notion of
$Q$-cohomology doesn't make sense unless we are talking about $V$-invariant
forms.

The simplification of this procedure is that the bosonic action now has a
potential term, so the vacua are localized at the minima of the potential.
Since the potential is essentially $\vert V\vert^2,$ the minima are at the
zeros of $V.$  To simplify the discussion, we will assume that we can choose a
holomorphic vector field with isolated zeros.  In fact, by a mathematical
theorem of Carrell and Lieberman, this property requires the Hodge numbers to
be diagonal \rCL.  These are just the manifolds in which we are interested from
the point of view of classification, as they have finite chiral rings.
Furthermore, such manifolds have the property that the $d_m$ cohomology is
isomorphic (as a vector space) to the ordinary de Rham cohomology.  Therefore,
no number of observables is lost in the addition of the potential to our
theory.

The physical classification of these theories rests on the calculation of
soliton numbers.  In the ordinary sigma model, the space of minimum bosonic
configurations is the entire target manifold (constant maps) and the solitons
are derived from a quantum-mechanical analysis.  Here things are much simpler:
the vacua are points, as in Landau-Ginzburg theories.  As usual, we consider an
infinite cylinder with compactified time.  Let us label the vacua (zeros of
$V$) $x_a.$  The solitons in the $ab$ sector correspond to time-independent
field configurations with $\phi(-\infty) = x_a$ and $\phi(+\infty) = x_b.$  The
solitons which saturate the Bogolmonyi bound minimize the energy functional.
We have, for (the bosonic part of) time-independent configurations,
$$\eqalign{E &= \int dx \left[
g_{i\oo{j}}\partial_x\phi^i\partial_x\phi^{\oo{j}} + m^2\partial_i
U\partial_{\oo{j}}Ug^{i\oo{j}}\right] \cr
&= \int dx[\partial_x\phi^i \pm m\partial^i U][\partial_x\phi^{\oo{j}} \pm
m\partial^{\oo{j}}U]g_{i\oo{j}} \mp m\int dx (\partial_x U),}$$
from which we derive the Bogolmonyi bound
$$E \geq m\vert U(\infty)-U(-\infty)\vert$$
($m$ can be incorporated into $U$ as well; henceforth we will put $m=1$).
This bound is saturated for trajectories for which
\eqn\traj{\partial_x\Phi = J\cdot V,}
where $J$ is the action of the complex structure.
Thus the solitons move along paths defined by the vector $JV,$ which is $V$
rotated by the complex structure tensor, i.e. $\Phi^\prime = JV$ (here we are
speaking of the real vector field $V+\oo{V}$).  We note here that any
trajectory obeying \traj, transformed by the flow defined by  $V,$ will still
obey \traj.  This follows because $U$ and the metric are invariant with respect
to $V.$

The subtleties of these theories are two-fold.  First, we need to know how to
compute the chiral ring and ensure that the $tt^*$ structure is left intact.
Secondly, we have a continuous set of classical soliton trajectories which
needs to be quantized by the method of collective coordinates.  When we perform
the quantization,\footnote{$^*$}{See, for example, Ref. \rRaj.  One extracts
the parameter describing the different solutions from the path integral.
Restoring the time dependence amounts to treating it as a quantum-mechanical
particle, so the space of instantons includes a one (or more) particle Hilbert
space, from which we will choose the state of lowest energy.} the collective
coordinate becomes a quantum-mechanical particle.  This technique is standard.
The first subtlety amounts to asking whether there is continuity at $m=0,$ i.e.
do the soliton numbers change discontinuously when we turn on this vector
field.  As we have discussed previously, the cohomologies are isomorphic, but
the theories may be different.  We may suddenly be describing a massive N=2
theory with a different classification -- after all, the configuration space
now only include $V$-invariant forms.  Without fully resolving these issues, we
will walk through a simple example, highlighting the general features.

Our example is just the sphere, ${\bf P}^1,$ endowed with its usual
(Fubini-Study) metric,
$$g_{z\zbar} = \del\oo{\del}\ln{(1+\vert z\vert^2)}.$$  The holomorphic vector
field is just a rotation of the sphere, say about the polar axis
($\varphi\rightarrow \varphi + c$).  The vector field $V$ has the form $$V =
iz\del_z - i\zbar\del_{\zbar}$$
and the function $U$ is $$U = -{\vert z\vert^2\over (1+\vert z\vert^2)}$$
(we sometimes use the notation $V$ for just the holomorphic piece, too), and so
$iV^z = z\del_z,$ which generates dilatations or lines of longitude emanating
from the north pole.  Specifically, the solitons are
$$z(t) = \rho\e{i\varphi_0}\e{mx},$$
where $\rho$ and $\varphi_0$ are arbitrary real parameters associated to
translation invariance and the $U(1)$ invariance generated by $V.$  The
translational symmetry means that solitons can appear with arbitrary momentum.
The minimal energy will have zero momentum.  The collective coordinate
$\varphi_0$ is simply the spatially constant part of the azimuthal angle
$\varphi$ and becomes a free particle upon quantization.  Explicitly, we define
$z = \rho\e{i\varphi}$ then single out the zero mode $\varphi =
\widetilde{\varphi}(x,t) + \varphi_0(t)$ and write the bosonic action as
$$\eqalign{S_{\rm bos} = &\phantom{+}\int dxdt {1\over (1+\rho^2)^2}\left[
\del_\mu\rho\del^\mu\rho +
\rho^2\del_\mu\widetilde{\varphi}\del^\mu\widetilde{\varphi} -
m^2\rho^2(1+\rho^2)\right]\cr & + \int dt \dot{\varphi_0}^2A +
\dot{\varphi_0}B.}$$
Here $A$ and $B$ are defined in terms of the other fields:
$$A =  \int dx {\rho^2\over (1+\rho^2)^2};\qquad B =  \int dx {\rho^2
\del_x\widetilde{\varphi}\over (1+\rho^2)^2}.$$ Expanding around the classical
soliton solution and performing the integration, we have $$A = {1\over 2m};
\qquad B = 0.$$  The action for $\varphi_0$ is thus a standard free single
particle quantum-mechanical action.

We know the full Hilbert space of a free particle in one (bound) dimension, and
its energy is minimized by the $n=0$ ground state.  This analysis would thus
tell us that there is just one Bogolomonyi soliton.  However, if we could
eliminate this state, the ground state(s) would appropriately be a doublet ($n
= \pm 1$).  This is indeed the correct representation under $U(1)$ induced by
the doublet of $SU(2).$  That is, the theory without potential -- what we are
interested in, after all -- has an $SU(2)$ symmetry and the solitons have been
shown (by independent analysis) to lie in the doublet of $SU(2).$  The
potential breaks the $SU(2)$ to $U(1),$ and the resulting solitons should thus
have charges $\pm \half$ under this $U(1).$   Unfortunately, to eliminate the
ground state we had impose the known solution.  A possible resolution of this
conundrum comes from the form of the N=2 algebra. in the models with potentials
from holomorphic Killing vectors.  If Ref. \rPot\ it was found that the algebra
contains central terms proportional to ${\cal L}_V.$  Thes terms are precisely
of the form as those which appear in the soliton sector of any N=2
supersymmetric theory.  For example, a Landau-Ginzburg theory with a
superpotential has central terms proportional to $\Delta W.$  Thus a non-zero
value of ${\cal L}_V = \del_\varphi$ (which is none other than the $U(1)$
charge) may be expected in the soliton sector.

\subsec{$\widetilde{\bf P}^2:$ A Good Test Case}

In this section we will present materials necessary for studying the manifold
$\widetilde{\bf P}^2,$ by which we mean the blow-up of the projective space
${\bf P}^2$ at a point (in section 3.4 this was denoted $\widetilde{\bf
P}_1^2).$  This manifold is particularly interesting for several reasons.  It
is a diagonal Fano variety with $c_1 > 0,$ so it defines a good quantum field
theory.  Further, it is not a coset space.  Coset spaces may prove to satisfy
the proposed link due to simplifications from a representation-theoretic
description.  However, this space has no simple treatment, so the equivalence
here would show that the link was more robust.  Another reason this space is
interesting is that though it is not simple, we do have some tools available to
help our treatment.  $\widetilde{\bf P}^2$ is the blow-up of a projective
space.  Due to this fact, we will be able to give an exceptional collection and
compute the bilinear form from the mathematical point of view.  Happily, this
space is also a toric variety.  These spaces were studied by Batyrev, who
showed that their quantum cohomology rings have a very simple description.
This knowledge is necessary for computing the soliton numbers through the
$tt^*$ equations in order to compare with the mathematical results.  What is
more, this space has holomorphic vector fields, since it is a toric variety.
Therefore, it may be treatable by the method of localization described in
section 4.2.  Finally, the space has only four cohomology classes, so the
calculations are not too messy.

Let us first describe the blow-up procedure. We begin by recalling the blow-up
of ${\bf C}^2$ at the origin, denoted $\widetilde{{\bf C}}^2.$  It is the
subset of ${\bf C}^2\times{\bf P}^1$ defined by
$${\bf C}^2 = \{(z_1,z_2;\lambda_1,\lambda_2) \in {\bf C}^2\times{\bf P}^1 :\>
z_1\lambda_2 - z_2\lambda_1 = 0 \}.$$
Thus the $\lambda$ part is determined to be the unique line through $(z_1,z_2)$
whenever $(z_1,z_2)\neq (0,0).$  At the origin, we have the entire ${\bf P}^1$
(all lines through $0$).  We call this ${\bf P}^1$ the exceptional divisor,
$E.$  Note that $\widetilde{{\bf C}}^2 - E \cong {\bf C}^2 - 0.$  To define the
blow-up of ${\bf P}^2,$ we choose a point $(0,0,1)$ and replace a neighborhood
isomorphic to ${\bf C}^2$ by its blow-up.  Thus,
$$\widetilde{{\bf P}}^2 = \{ (\mu_1,\mu_2,\mu_3; \lambda_1,\lambda_2) \in {\bf
P}^2\times {\bf P}^1 :\> \mu_1\lambda_2 - \mu_2\lambda_2 = 0\}.$$
This is the zero locus of a homogeneous function of bi-degree $(1,1)$ in ${\bf
P}^2\times{\bf P}^1.$

We can also understand this space as a toric variety, being one of the
rational, ruled, or Hirzebruch, surfaces.  The study of toric manifolds is
quite a broad subject.  A readable hands-on introduction is given by Batyrev in
his paper on quantum rings of toric varieties, the results of which we shall
use presently \rBat.

In the language of toric varieties, the space $\widetilde{\bf P}^2$ is
described by the diagram (drawn on a plane) \toricdiag, which is interpreted as
follows.  We define a complex variable for each arrow, begining with the space
${\bf C}^4 - V.$  Here $V$ is an open set containing points which must be
removed so that the group action by which we mod out has no fixed points.  From
\toricdiag\ we get that\footnote{$^*$}{This set is defined by taking the union
of all sets obtained by setting to zero all coordinates corresponding to
vectors of a primitive collection.  A primitive collection consists of vectors
not generating a single cone (there are four two-dimensional and four
one-dimensional cones in \toricdiag), though any subset of those vectors
generates a single cone of the diagram.}
$$V = \{ z_1 = z_2 = 0\}\cup\{z_3=z_4= 0\}.$$
We then mod out by the action of $({\bf C}^*)^2,$ which is derived from the
independent relations among the arrows:  $\vec{v}_1 + \vec{v}_2 = 0,$
$\vec{v}_1 + \vec{v}_3 + \vec{v}_4 = 0.$  Thus we act without fixed point by
$(\lambda,\rho)\in ({\bf C}^*)^2$ on ${\bf C}^2-V$ by sending
\eqn\toract{(z_1,z_2,z_3,z_4)\mapsto (\rho\lambda z_1, \lambda z_2,\rho z_3,
\rho z_4).}
We can see here that $(z_1,z_2,z_3,z_4)\rightarrow(z_3,z_4)\in{\bf P}^1$
represents a fibering over ${\bf P}^1$ equal to the projectivization of the
bundle $\O\oplus\O(1),$ as the $\lambda$ action represents the projective
equivalence on the fiber and the $\rho$ action on $z_1$ denotes that it is a
coordinate of an $\O(1)$ bundle.  The $n^{\rm th}$ Hirzebruch surface, $H_n,$
gives $\vec{v}_4$ a height of $n$ (instead of one) and is equal to the total
space of the bundle ${\bf P}(\O\oplus\O(n)).$

In order to compute the $tt^*$ equations for this manifold, we need the chiral
ring coefficients $C_{ij}{}^k.$  In fact these can be computed in the purely
topological theory, since $C_{ijk} = \langle \phi_i\phi_j\phi_k\rangle,$ and
the indices are raised with the topological metric $\eta_{ij}.$  The
topological correlation functions are obtained by passing to the topological
limit, in which the path integral becomes an integral over the moduli space of
instantons (not to be confused with parameter space).  The instantons are
holomorphic maps.  The topological observables, as we have discussed,
correspond to cohomology classes of forms, and can be chosen to have support on
their Poincar\'e dual cycles, $L_{i}.$  The correlators
$\langle\phi_i(p_i)\phi_j(p_j)\phi_k(p_k)\rangle$ just count the number of
holomorphic maps taking $p_i$ to $L_i$ (when that number is
finite\footnote{$^{**}$}{If the formal dimension of such maps is zero but there
is a continuous family, then the ``number'' of maps is replaced by the Euler
characteristic of a vector bundle over this family.}), weighted by $\exp(-dA),$
where $d$ is the degree of the instanton and $A$ is the area of the image
(which depends on the K\"ahler class of the target space).

Batyrev has calculated the ring coefficients for toric varieties. We quote his
results without proof for the $n^{\rm th}$ Hirzebruch surface, $H_n,$ obtained
by taking $\vec{v}_4 = (-1,n)$ (our example is $n = 1$) and generalizing the
action of $({\bf C}^*)^2$ in \toract\ by changing the $\rho\lambda z_1$ to
$\rho^n\lambda z_1$ on the right hand side.  For this space, Batyrev's
prescription gives a ring with two generators, $z_1$ and $z_2,$ and the
following relations:
$$\eqalign{z_1^2 &= \e{-\beta}z_2^n \cr
z_2^2 &= \e{-\alpha} - nz_1z_2.}$$
In the abvoe, the $\alpha$ and $\beta$ are parameters describing the K\"ahler
class.  As they represent the areas of the two homology cycles, they should
both be positive.  We note here that in the large radius limit, where the Ricci
curvature goes to zero and these areas go to infinity, we recover the ordinary
cohomology ring or intersection of $H_n.$

The ring gives us the values $C_{ij}^{k}.$  To compute all the correlators, we
need $C_{ijk}$ which we can get by lowering indices with $\eta_{ij} = C_{ij0} =
C_{ij}{}^m\eta_{m0}$ (where the subscript $0$ represents the identity element
of the ring).  But $\eta_{m0}$ represents the one-point function, which is
determined by the anomalous charge conservation.  That is, the chiral charge
(i.e. the form degree) is violated by $2d$ (the dimension of the space) units,
so only the top form ($z_1z_2$ here) can have nonzero one-point function.  We
can normalize this to be unity.  Thus we have all the information necessary to
write down the full $tt^*$ equations.

The $tt^*$ equations will be nonlinear differential equations for the metric
$g_{i\oo{j}}.$  Since this is a matrix, they are coupled differential equations
representing the different entries.  It is almost certain that they cannot be
solved analytically by today's methods.  One might hope to obtain the soliton
numbers through a numerical analysis.  A similar analysis was performed in
\rWLH\ in a computation involving Landau-Ginzburg theories.  In that paper, the
author started from the conformal (homogeneous) point and iterated out to the
infrared limit.  To do so, one needs the values and first derivatives of the
metric at the conformal point.  In the Landau-Ginzburg case, explicit formulas
for these values provide necessary ingredients.  It was found that convergence
of the solution also determines the boundary conditions.  This is consistent
with the cases which have been solved analytically.  For the topological sigma
model, certain asymptotic expressions for the metric are known (see section 5
of Ref. \rCVNEW).  One expects that these data and regularity of the solutions
would once again determine the boundary conditions needed to proceed with a
numerical computation of the soliton numbers from the physical viewpoint.

We should mention here that toric varieties all have holomorphic actions by
vector fields (and one can write metrics invariant with respect to such
actions, so that they are holomorphic Killing vectors).  A toric variety is
constructed from some open set in ${\bf C}^n$ by modding out by $({\bf
C}^*)^r.$  Roughly speaking, this leaves us with at least $({\bf C}^*)^{n-r}$
independent ${\bf C}^*$ actions of the form $z_j \mapsto \lambda z_j.$  In our
example, two remaining actions are
$$(z_1,z_2,z_3,z_4) \mapsto (sz_1,z_2,tz_3,z_4), \qquad s,t\in {\bf C}^*.$$
Let us consider the $S^1$ action defined by setting $s=t^2\in U(1).$  The
vector field which generates this action has four isolated fixed points, in
one-to-one correspondence with the number of vacua or cohomology elements for
this space.  They are $(0,1,0,1),(0,1,1,0),(1,0,0,1),$ and $(1,0,1,0).$  This
space is clearly able to be analyzed by the method of localization.  However,
it will certainly be necessary to clean up the ${\bf P}^1$ case before
proceeding in this direction.

{}From the mathematical point of view, exceptional collections over
$\widetilde{\bf P}^2,$ as well as the other rational, ruled surfaces, were
studied by Kvichansky and Nogin in \rHelices, and by Nogin in \rNogin.  They
found exceptional collections, as well as those which generate helices, in
which shifting a sheaf right $n$ times corresponds to tensoring by the
canonical bundle of $\widetilde{\bf P}^2.$  In fact, these authors found
foundations consisting of line bundles.  We describe line bundles using the
equivalence of divisors and line bundles discussed for ${\bf P}^n$ in
3.2\footnote{$^*$}{For other varieties the hyperplane bundle represents the
pull-back bundle under an imbedding into projective space.}.  The two homology
cycles correspond to one on ${\bf P}^2$ and the exceptional divisor.  Denote
these by $F$ and $C,$ respectively; products of their corresponding line
bundles $L_F$ and $L_C$ will be denoted $(L_F)^a\otimes (L_C)^b = aF + bC =
(a,b).$  The authors showed that the collections $\{ \O, \O(1,k), \O(1,k+2),
\O(2,2) \}$ are helices for all $k.$  It would be very interesting to use these
data to compute the ``mathematical'' soliton numbers $\chi(E_i,E_j)$ and
compare with results derived from physics.

\newsec{Prolegomena to Any Future Math-Physics}
We have detailed an interesting open problem in mathematics and physics, along
with several proposals for establishing a link between classification of N=2
theories and helices of exceptional sheaves.  Currently, there is no definite
connection, though the two areas have been shown to be related through
examples.  Further, a categorical link to the algebras of quivers has been
discussed.

Clearly, there are many approaches to solving this problem and much work needs
to be done in all directions.  One would like to amass more ``experimental''
evidence through a detailed exploration of a wide range of examples.
Physicists would like the mathematical theory to be more mature, in order to
develop better intuition for why solitons could have such an abstruse origin.
The situation is much like the status of supersymmetry and de Rham theory and
Morse theory, before Witten's famous papers relating the two.  Many roads can
be taken; is one likely to lead to such a fruitful discovery?  In the absence
of hard facts, we have amassed circumstantial evidence that some bridge between
the theories should exist.  The author trusts the reader to be well-skilled to
investigate this problem, and invites him/her to establish this elusive
connection.

\bigbreak\bigskip\bigskip\centerline{{\bf Acknowledgements}}\nobreak
Much of this paper is a compilation of the work of others who have helped me
with patient explanations and keen observations.  Principal among these are C.
Vafa and M. M. Kapranov, to whom I am most grateful.  I also thank S. Cecotti,
K. Intriligator, A. E. Polischuk, and V. Sadov for conversations and
communications.  This work was supported in part by Fannie and John Hertz
Foundation and by NSF contract
PHY-92-8167.

\listrefs
\listfigs

\end